\documentclass[structabstract]{aa}  
\pdfoutput=1 
\usepackage{natbib, graphicx}
\usepackage{amsmath}
\usepackage[flushleft]{threeparttable}
\usepackage{subfigure}
\usepackage{color}
\usepackage{txfonts}
\usepackage{multirow}
\usepackage{hyperref}
\usepackage{booktabs}
\usepackage{pdflscape}
\usepackage{adjustbox}
\usepackage{caption}
\usepackage{ftnxtra}
\usepackage{orcidlink}
\usepackage{fnpos} 
\usepackage[
    starfontserif 
    ]{starfont}
\DeclareSymbolFont{starfontsym}{OT1}{sts}{m}{n}
\DeclareMathSymbol{\mathJupiter}{\mathord}{starfontsym}{106}

\newcommand{\teff}{T$_\mathrm{eff}$}
\newcommand{\logg}{$\log{g}$}
\newcommand{\feh}{[Fe/H]}
\newcommand{\vt}{$\xi$}
\newcommand{\vmac}{v$_{\rm mac}$}
\newcommand{\Rp}{R$_P$}
\newcommand{\Mp}{M$_P$}
\newcommand{\Rj}{R$_J$}
\newcommand{\Mj}{M$_J$}
\newcommand{\Rstar}{R$_\star$}
\newcommand{\Mstar}{M$_\star$}

\newcommand{\Msun}{M$_\odot$}
\newcommand{\kms}{km\,s$^{-1}$}
\newcommand{\vsini}{v\ $\sin{i}$}

\DeclareCaptionLabelFormat{continued}{#1~#2~Continued}
\begin{document}

   \title{Ariel stellar characterisation}
   \subtitle{III. Fast rotators and new FGK stars in the Ariel Mission Candidate Sample\thanks{Based on data from public telescope archives and from observations collected at the ESO under programmes: 109.23J9, 110.24BU, and 111.2542 (PI: C. Danielski), at the Large Binocular Telescope Observatory under the programme 2021\_2022\_25 (PI: M. Rainer), with the Italian Telescopio Nazionale Galileo under programmes AOT41\_TAC25 (PI: S. Benatti) and AOT46\_TAC27 (PI: M. Rainer), and with the Southern African Large Telescope under the programme 2023-1-SCI-005 (PI: He{\l}miniak).}}
   \author{
        Tsantaki, M. \inst{\ref{oaa}}\orcidlink{0000-0002-0552-2313}
        \and    
        Magrini, L. \inst{\ref{oaa}}\orcidlink{0000-0003-4486-6802}  
        \and
        Danielski, C. \inst{\ref{oaa}}\orcidlink{0000-0002-3729-2663} 
        \and
        Bossini, D. \inst{\ref{uop}, \ref{oapd}}\orcidlink{0000-0002-9480-8400}
        \and
        Turrini, D. \inst{\ref{oato}, \ref{iaps}}
        \and
        Moedas, N. \inst{\ref{porto}}
        \and
        Folsom, C. P. \inst{\ref{tartu}}
        \and
        Ramler, H. \inst{\ref{tartu}}
        \and 
        Biazzo, K. \inst{\ref{oar}} 
        \and
        Campante, T. L. \inst{\ref{porto}, \ref{fcup}}\orcidlink{0000-0002-4588-5389}
        \and 
        Delgado-Mena, E. \inst{\ref{cab}, \ref{porto}}
        \and
        da Silva, R. \inst{\ref{oar}, \ref{asi}}\orcidlink{0000-0003-4788-677X}
        \and
        Sousa, S. G.\inst{\ref{porto}}        
        \and 
        Benatti, S. \inst{\ref{oap}}
        \and 
        Casali, G. \inst{\ref{anu}, \ref{astro3d}, \ref{oab}}
        \and 
        He{\l}miniak, K. G. \inst{\ref{nicolaus}}\orcidlink{0000-0002-7650-3603}
        \and 
        Rainer, M. \inst{\ref{oabr}}
        \and
        Sanna, N. \inst{\ref{oaa}}
    }
    \institute{
    INAF -- Osservatorio Astrofisico di Arcetri, Largo E. Fermi 5, 50125 Firenze, Italy\label{oaa}
    \and
    Department of Physics and Astronomy G. Galilei, University of Padova, Vicolo dell'Osservatorio 3, I-35122, Padova, Italy\label{uop}
    \and
    INAF -- Osservatorio Astronomico di Padova, Vicolo dell'Osservatorio 5, 35122, Padova, Italy \label{oapd}
    \and
    INAF -- Osservatorio Astrofisico di Torino, via Osservatorio 20, 10025, Pino Torinese\label{oato}
    \and
    INAF -- IAPS, Via Fosso del Cavaliere 100, 00133, Rome, Italy\label{iaps}
    \and
    Tartu Observatory, University of Tartu, Observatooriumi 1, Tõravere, 61602 Tartumaa, Estonia \label{tartu}
    \and
    INAF -- Osservatorio Astronomico di Roma, Via Frascati 33, 00040 Monte Porzio Catone (RM), Italy\label{oar}
    \and
    Instituto de Astrofísica e Ciências do Espaço, Universidade do Porto, CAUP, Rua das Estrelas, 4150-762 Porto, Portugal \label{porto}
    \and 
    Departamento de F\'{\i}sica e Astronomia, Faculdade de Ci\^{e}ncias da Universidade do Porto, Rua do Campo Alegre, s/n, 4169-007 Porto, Portugal\label{fcup}
    \and
    Centro de Astrobiolog\'ia (CAB), CSIC-INTA, Camino Bajo del Castillo s/n, 28692, Villanueva de la Ca\~nada (Madrid), Spain \label{cab}
    \and
    Agenzia Spaziale Italiana, Space Science Data Center, via del Politecnico snc, 00133 Rome, Italy\label{asi}
    \and 
    INAF -- Osservatorio Astronomico di Palermo, Piazza del Parlamento, 1, 90134 Palermo, Italy\label{oap}
   \and
   Research School of Astronomy \& Astrophysics, Australian National University, Cotter Rd., Weston, ACT 2611, Australia \label{anu}
   \and
   ARC Centre of Excellence for All Sky Astrophysics in 3 Dimensions (ASTRO 3D), Stromlo, Australia \label{astro3d}
   \and
   INAF -- Osservatorio di Astrofisica e Scienza dello Spazio di Bologna, via P. Gobetti 93/3, 40129, Bologna, Italy  \label{oab}
   \and
   Nicolaus Copernicus Astronomical Center, Polish Academy of Sciences, ul. Rabia\'{n}ska 8, 87-100 Toru\'{n}, Poland \label{nicolaus}
   \and
   INAF -- Osservatorio Astronomico di Brera, Via E. Bianchi 46, 23807 Merate (LC), Italy \label{oabr}
    }

   \date{Received 18 November 2024; accepted XXXX}
   \authorrunning{M. Tsantaki}
   \titlerunning{Ariel stellar characterisation: fast rotators and new sample}
   \abstract
{The next mission dedicated to the study of planetary atmospheres is the Ariel space mission, planned for launch in 2029, which will observe a variety of planetary systems belonging to different classes around stars with spectral types from M to A. To optimise the scientific outcome of the mission, such stars need to be homogeneously characterised beforehand.}
{In this work, we focus on a methodology based on spectral synthesis for the characterisation of FGK-type stars from the Ariel Tier 1 Mission Candidate Sample (MCS) which exhibit fast rotation. In addition, we analyse 108 slow-rotating FGK-type stars, with either new observations or archival spectra available, consistently as in our previous work using the equivalent width (EW) analysis.}
{To ensure consistency between our methods, we re-analysed a sample of FGK-type stars with the spectral synthesis method and compared it to our previous work. The results of our analysis show excellent agreement with the previous set of derived parameters. }
{We provide homogeneous effective temperature, surface gravity, metallicity, projected rotational velocity, and stellar mass for a sample of 36 fast rotators with the spectral synthesis technique, and we include 108 FGK-type dwarfs with the EW analysis. An additional 25 stars were analysed with the spectral synthesis method because their EW analysis did not converge on the final parameters. We computed their orbital parameters establishing whether they belong to the {Galactic} thin or thick discs. With the current set of stellar parameters, we almost double the analysed hosts in the Ariel MCS to 353 stars in total.}
{Using our homogeneous set of stellar parameters, we studied the correlations between stellar and planetary properties for the Ariel MCS analysed so far. We confirmed a close relationship between stellar mass (up to 1.8\,\Msun) and giant planet radius, with more inflated planets at lower metallicity. We confirm that giant planets are more frequent around more metal-rich stars that belong to the thin disc, while lower-mass planets are also found in more metal-poor environments, and are more frequent than giant planets in the thick disc as also seen in other works in the literature.}

\maketitle

\section{Introduction}\label{sec:intro}

In 2020, the European Space Agency (ESA) adopted the Ariel medium class space mission as a large program fully dedicated to the chemical characterisation of the atmospheres of planets beyond the Solar System. After its scheduled launch in 2029, Ariel will survey a thousand exoplanets, spanning a variety of properties, to perform a broad study of exoplanetary atmospheres \citep{Tinetti2018}. The ultimate goal of the mission is to understand the physical processes behind both planetary formation and atmospheric evolution for the different classes of planets \citep{Turrini2021, tinetti2021}. More specifically, Ariel will reveal chemical fingerprints of gases and condensates in the planetary atmospheres, including their thermal structure and elemental composition, enabling the investigation of all corners of the exoplanet population, from temperate terrestrial to ultra-hot Jupiters. Meeting such an objective requires performing a population study with a homogeneous approach, to allow for a direct comparison between all planetary systems that will be observed. In particular, for correctly interpreting the atmospheric data that Ariel will retrieve, each host star in the sample must be characterised with high precision in a uniform way \citep{Danielski2022}. A sample of uniform and self-consistent stellar parameters (e.g., stellar atmospheric and orbital parameters, as well as chemical composition, mass, age, and stellar activity) will establish a robust reference frame that will enable us to perform comparative planetary studies for the thousand planets, and hence to shed light on their formation and evolution on a global scale.

The precise and accurate characterisation of planet-host stars (PHS) is therefore crucial for the success of the science goals of the Ariel mission. In this context, a methodology has been compiled to analyse the FGK-type stars in \citet[][hereafter M22]{Magrini2022}. The spectral analysis in M22 is based on the measurements of equivalent widths (EWs) of iron neutral and ionised lines in spectra obtained in high resolution and with high signal-to-noise ratio (S/N). The method in M22 uses additional constraints from photometric and astrometric data obtained from 2MASS \citep{Skrutskie2006} and {\it Gaia} DR3 \citep{Gaia2023} to determine the surface gravity, which is set fixed to its spectro-photometric value which is usually defined in the literature as trigonometric surface gravity \citep[e.g.,][]{mortier13_transits, tsantaki13}. 

This method has been successfully applied to FGK-type targets that are restricted to low projected rotational velocities (\vsini). An intrinsic limitation of this method is, indeed, stellar rotation. High rotational velocities cause spectral lines to be broadened and shallower, and thus blended, which in turn makes their EWs extremely difficult to be measured. The limiting \vsini\, value of this method depends on the spectral type and usually we can measure reliable EWs for stars with \vsini\, empirically set up to 10-15\,\kms~ \citep[e.g.,][]{tsantaki14}. For main sequence stars, we expect the hotter stars (starting from early F-type) to rotate faster due to their more extended radiative zones in their outermost layers compared to their cooler counterparts where convection dominates \citep[e.g.,][]{Barnes2003}.

Spectral synthesis has been proven an excellent alternative method to derive stellar parameters (namely effective temperature, \teff, surface gravity, \logg, and iron metallicity, \feh) of stars with significant rotation with various methodologies available in the literature \cite[e.g.,][]{Blanco-Cuaresma2014, Piskunov2017, Tsantaki2018, Tabernero2022}. The principle of the spectral synthesis technique relies on matching a model spectrum synthesised for a set of stellar parameters and for a specific wavelength region with an observed spectrum. The best-fit parameters are obtained after a minimisation process. 

In the present paper, we focus on describing a specific methodology based on the spectral synthesis technique to derive stellar parameters for the fast rotators (up to a \teff$\sim$7300\,K and \vsini$\sim$70\,\kms~ in this sample). We used the sample of M22 to verify our spectral synthesis methodology and to guarantee uniformity with the previous work. We then applied the spectral synthesis technique to fast rotating stars with planets aimed to be observed by the Ariel space mission and for which we have high-resolution spectra either from archival data or from our new observations. Finally, we analysed additional new and archival data for slow rotating FGK-type stars as a continuation of the work of M22 with the aim at increasing the number of stars with planets to be observed by the Ariel mission. Future works will be devoted to the analysis of stars of different spectral types, such as M dwarfs (Maldonado et al., in prep) and very hot stars with spectral type $\geq$A (Ramler et al., in prep.). The results of the present work significantly increase the parameter space of characterised PHS, in particular including for the first time early F stars which lay at the more massive end. Finally, with the results of the present work, the number of stars with planets to be observed by the Ariel mission with homogeneous parameters has increased by more than $\sim$\,48\% in size allowing correlations between stellar and planetary properties to be better studied. 

This paper is organized as follows. In Sect.~\ref{sec:sample}, we present the sample of high-resolution spectroscopic data used for our analysis. In Sect.~\ref{sec:method}, we describe the spectroscopic methods we used for the analysis, presenting a validation on the homogeneity with our past work. In Sect.~\ref{sec:results}, we present the results of the spectroscopic analysis and kinematic properties of the new samples. In Sect.~\ref{sec:planets}, we discuss the relationships between the PHS and the properties of their planetary companions. We provide a summary and conclusions of this work in Sect.~\ref{sec:summary}. 

\section{The data sample}\label{sec:sample}

The stars presented in this work are part of the Ariel Mission Candidate Sample (MCS), as of 2024, which contains exoplanets suitable for Ariel observations \citep{Edwards2019, Edwards2022}. However, the Ariel MCS will continue to be dynamic as new exoplanets are discovered and the optimal observing strategy of the mission is defined. The final target list will contain up to 1000 exoplanets in both single and multiple systems and will be endorsed by the Ariel Science Team and reviewed under the ESA Advisory Structure before launch. The Ariel ``Stellar Characterisation'' working group is responsible for characterising all PHS in the Ariel MCS with the precision and accuracy needed to fulfil the goals of the mission. To achieve this, we have employed an ongoing ground-based monitoring campaign in both hemispheres using facilities with high-resolution spectrographs for more than six years. In addition, public archives are routinely searched for available high-resolution and high S/N data.

The first set of data were analysed in M22 while the spectra analysed in the present work are gathered either from archival databases or from new observations obtained by our team focusing only on FGK-type stars. We also acquired new observations for stars whose archival spectra were not available or, if they were, their quality was not sufficient for spectral characterisation at the requested precision level. The spectroscopic analysis of this work and the following ones to cover all PHS of the Ariel MCS will be included in an incremental way into a much needed homogeneous and self-coherent catalogue to provide a variety of stellar parameters (such as atmospheric parameters, elemental abundances, activity indicators, ages, masses, radii) that will be made publicly available at \url{https://sites.google.com/inaf.it/arielstellarcatalogue}{}.

\subsection{New observations}

We include observations from three successful observing runs at the European Southern Observatory (ESO) facilities with the Ultraviolet and Visual Echelle Spectrograph (UVES) at the Very Large Telescope \citep[VLT,][]{Dekker2000}. The spectra were obtained with a resolving power R$\approx$60\,000, covering a spectral range that includes the blue (326-454\,nm) and the red (476-684\,nm) wavelength regions with a goal of S/N higher than 120 at 600\,nm. We obtained new UVES spectra for 24 stars.

The data set is increased with spectra obtained by the High Accuracy Radial velocity Planet Searcher in the northern hemisphere (HARPS-N) at the Telescopio Nazionale Galileo \citep[TNG,][]{Cosentino2012}. The largest part of the observations have already been analysed in M22 while in the present work, we include new spectra for four stars: KELT-1, TOI-1601, TOI-1333, and TOI-628. The resolving power of HARPS-N is R$\approx$115\,000 and the wavelength region covered is 380-690\,nm. 

New spectra of four Ariel PHS were also collected using the high-resolution spectrograph (HRS) mounted on the Southern African Large Telescope (SALT) situated at the South African Astronomical Observatory at Sutherland (South Africa). The HRS covers a wavelength range of 370-890\,nm with a resolving power R$\approx$65\,000 \citep{Bramall2012}. The SALT/HRS has the advantage of including the region of the oxygen triplet at 777.7\,nm and in this context two stars (WASP-79 and WASP-94A) were re-observed from this facility even though their archival spectra were analysed in M22.

Finally, spectra of two stars (Kepler-1517 and HAT-P-65) were obtained with the Potsdam Echelle Polarimetric and Spectroscopic Instrument (PEPSI) at the 8.4\,m Large Binocular Telescope \citep[LBT,][]{Strassmeier2003}. The PEPSI observations were performed for the fainter northern targets with a resolving power R$\approx$50\,000 and covering the spectral region 383-912\,nm.

All spectra were reduced by the standard pipelines available for each instrument \citep[][respectively]{Ballester2000, Cosentino2014, Crawford2010, Strassmeier2018} and their multiple exposures were co-added together to increase the S/N after they were adjusted for radial velocity (RV) shifts. In total, we have 32 newly collected spectra. 

\subsection{Archival data}

We cross-matched the PHS in the Ariel MCS with the public spectral archive of various high-resolution spectrographs. We include in our sample 137 additional FGK-type stars with high enough S/N to satisfy the requirements for precise parameter and chemical abundance determinations. Most of the archival spectra have S/N higher than 100 and come from the ESO and TNG archives, mainly from UVES and HARPS (N and S) spectrographs. Fewer spectra were also obtained from VLT/ESPRESSO, 2.2\,m of ESO/FEROS, CFHT/ESPADONS, 1.93~m of OHP/SOPHIE, NOT/FIES, LBT/PEPSI, and TBL/NARVAL (see M22 for more details). The final list of stars analysed in this work is presented in Table~\ref{sample_params}. 

\section{Methodology of the spectral analysis}\label{sec:method}

The methodology adopted to provide stellar parameters for the FGK-type PHS in the Ariel MCS is presented in detail in M22 and briefly summarised in Sect.~\ref{sec:ew_analysis}. This method has been successfully applied to characterise 187 FGK-type stars with low rotation (\vsini\,$\lesssim$\,10-15\,\kms). In the present work, we apply the same analysis to an additional 108 FGK-type stars which were added to increase the number of stars homogeneously characterised in the Ariel MCS. However, as we mentioned before, due to line blending, the EW analysis cannot be applied to stars with high rotation or very broad lines, which affects mostly hot stars. For this reason, we were not able to measure correctly the EWs for 61 stars and used the spectral synthesis technique instead which is summarised in Sect.~\ref{sec:fasma}.

Since homogeneity for stellar parameters of the PHS in the Ariel MCS is a key aspect, we focus on ensuring that both methods are on the same scale. In order to guarantee homogeneity of our results obtained with the two different methods, we re-analysed the sample of M22 with the spectral synthesis technique proving that parameters derived with the two methods are consistent. In this Section, we summarise both methodologies and provide a comparison of their results to validate the agreement between the two methods. 

\subsection{The EW analysis of M22}\label{sec:ew_analysis}

The spectral analysis in M22 is based on the measurements of EWs of iron neutral and ionised lines. In the spectral analysis with EWs, ionization balance and excitation equilibrium are imposed on the iron abundances to obtain \teff~ and \logg. The microturbulent velocity (\vt) is instead obtained from the non-dependence of the iron abundance A($\ion{Fe}{I}$)\footnote{A($\ion{Fe}{I}$)=12+log $\frac{N(\ion{Fe}{I})}{N(H)}$, where N($\ion{Fe}{I}$) and N(H) are the number of atoms for the $\ion{Fe}{I}$ and H respectively.} to the reduced EWs (defined as the EW divided by the wavelength of the line). The final iron metallicity is obtained from the average abundances of the neutral iron lines. However, the method in M22 also uses additional constraints from 2MASS photometry and {\it Gaia} DR3 \citep{Gaia2023} photometric and astrometric data to determine the surface gravity. The adopted method is thus iterative, starting from parameters that are initially obtained from purely spectral analysis. Using this first set of parameters, the stellar mass is determined by isochrone fitting and subsequently the surface gravity is derived (see Sect.~\ref{sec:fasma}). By fixing the \logg~ to its trigonometric value, the spectral analysis is repeated, obtaining a new set of stellar parameters and masses. The procedure is repeated for two runs. The selection of trigonometric gravity over the spectroscopic was justified and discussed in \cite{brucalassi22} by presenting several methods of spectral analysis and comparing among them.

The adopted line list was developed for the {\em Gaia}-ESO survey \citep[GES,][]{Gilmore2022, Randich2022} and presented in \cite{Heiter2021}. The radiative transfer code used for the analysis was MOOG\footnote{\url{https://www.as.utexas.edu/~chris/moog.html}} \citep{sneden1973} which is automatised with the wrapper FAMA \citep{Magrini2013} to compute stellar parameters and chemical abundances. The MARCS model atmospheres \citep{gustafsson08} are used in both the spherical (giant stars) and plane-parallel (dwarf stars) geometries. Although the M22 method is fast and precise for FGK-type stars, it begins to fail when stellar rotation renders EWs useless, i.e. \vsini\,$\gtrsim$\,15\,\kms. We note that the sample of M22 contains some cases with \vsini\,$\gtrsim$\,15\,\kms. We re-analysed them in Sect.~\ref{sec:results} to inspect any discrepancies. 

\begin{figure}
  \centering
  \includegraphics[width=1.0\linewidth]{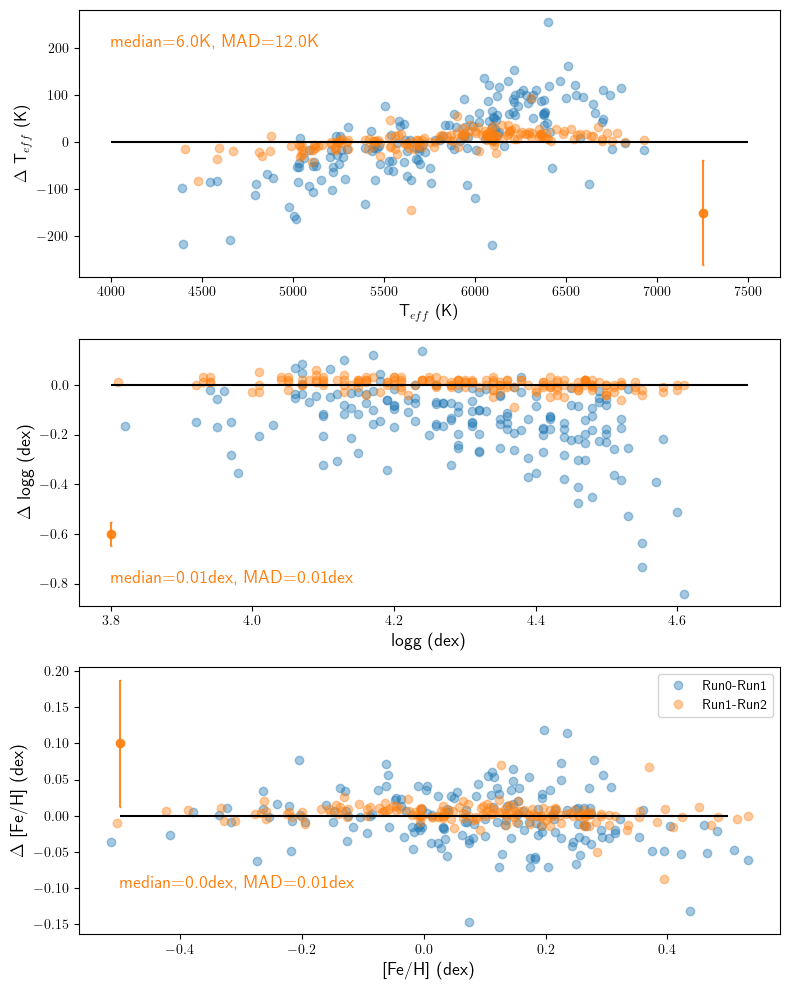}  \\ 
  \caption{Comparisons of the parameters derived with spectral synthesis in the different runs. The panels show the differences between Run 0, Run 1, and Run 2 for \teff~ (upper panel), \logg~ (central panel) and \feh~ (lower panel). The median differences and the MAD of the last runs are indicated in orange and the black lines represent the zero difference. The orange points show the average error bars of the differences in the y-axes.}
  \label{comparison_runs}
\end{figure}

\begin{figure}
  \centering
  \includegraphics[width=1.0\linewidth]{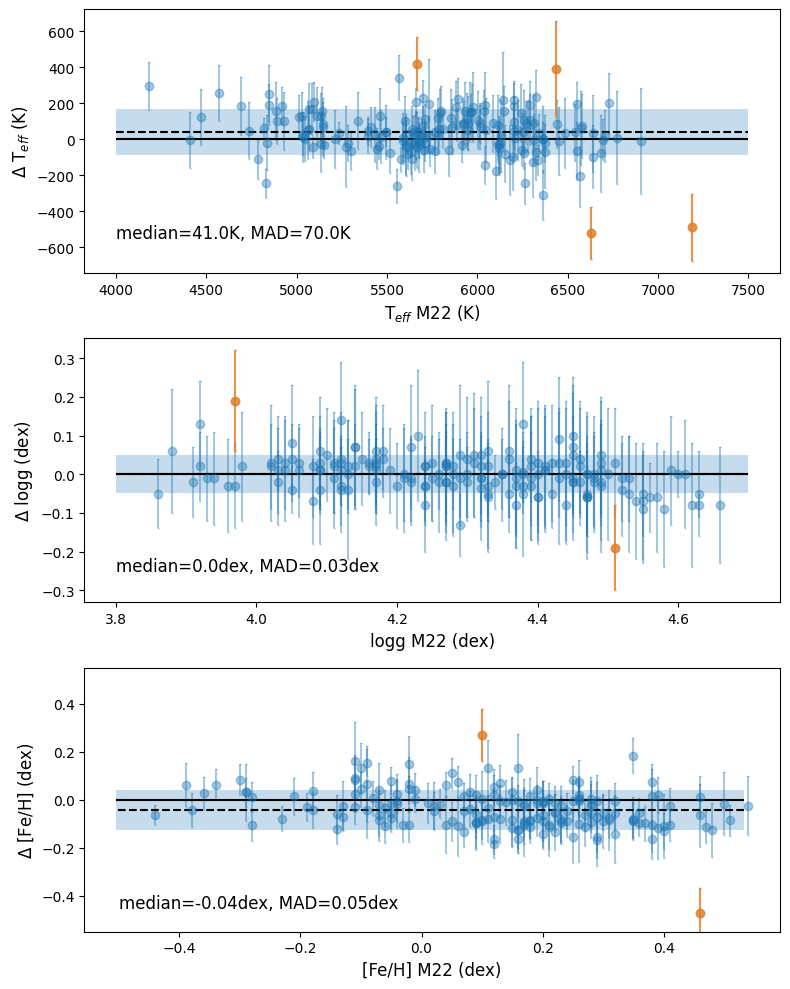}  \\ 
  \caption{Results of our parameters from spectral synthesis in comparison with those of M22. The panels show the differences (this work--M22) in \teff~ (upper panel), \logg$_{\rm trig}$ (central panel) and \feh~ (lower panel) as a function of the respectively parameters derived in M22. The orange points in all panels indicate stars with differences higher than 3\,$\sigma$. The shaded areas correspond to the 1\,$\sigma$ of the differences from the median. The dashed lines show the median differences for each parameter while the solid lines are zero.}
  \label{comparison_3rdrun}
\end{figure}

\begin{figure*}
  \centering
  \includegraphics[width=1.0\linewidth]{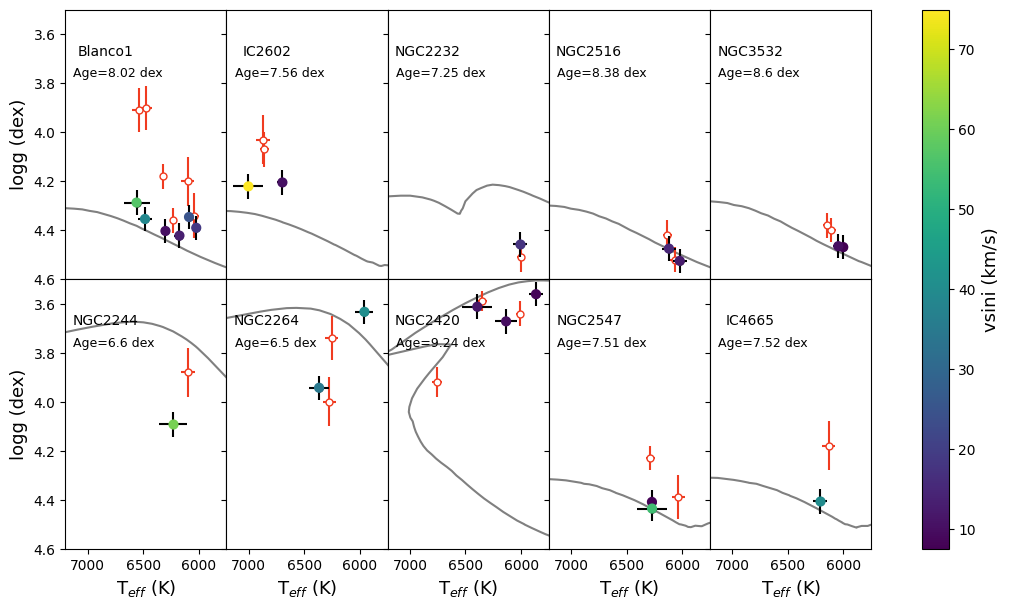}  \\ 
  \caption{Kiel diagrams for a sample of stars in clusters observed by the GES. The parameters derived from our methodology are colour coded to \vsini~ and the GES parameters are represented by red open circles. The gray lines are the PARSEC isochrones calculated with the ages and metallicities derived by the GES \citep{Jackson2022, Randich2022}.}
  \label{cluster}
\end{figure*}

\subsection{Spectral synthesis with fixed surface gravity}\label{sec:fasma}

Our goal is to follow as close as possible to the methodology described above, using the spectral synthesis technique instead of the EW method. For the spectral analysis, we used the FASMA\footnote{\url{https://github.com/MariaTsantaki/FASMA-synthesis}} software \citep{Tsantaki2018, Tsantaki2020}, which is based on MOOG (version 2019) and is the same radiative transfer code as in M22. FASMA creates synthetic spectra on-the-fly to deliver the best-fit parameters after a non-linear least-squares fit (Levenberg-Marquardt algorithm). The line list adopted in this work is based on the one compiled in \cite{Tsantaki2018} which is mostly comprised of iron lines but we used the atomic line data, i.e. the oscillator strengths, for the lines in common from the line list of the {\em Gaia}-ESO survey \citep{Heiter2021} which was also used in M22. For the rest of the lines, we kept the atomic data of the original line list which were either from the VALD database \citep{Ryabchikova2015} or calibrated from observed spectra (see \cite{Tsantaki2018} for details). The damping parameters are based on the ABO theory \citep{Barklem2000} when available, or in any other case, we used the Unsold approximation \citep{unsold1955}. We used the standard solar abundances from \cite{asplund09} internally adopted by MOOG, except for iron, for which we adopted A(Fe) = 7.45\,dex as in M22. The uncertainties on the stellar parameters are derived from the covariance matrix constructed by the non-linear least-squares fit. The methodology to compute the stellar parameters is a three-step process closely following M22. 

We summarise the steps of our iterative method below in the three runs: 
\begin{itemize}
\item[R.0] A spectral analysis is deployed to derive the first set of stellar parameters (\teff$_{, 0}$, \logg$_{0}$, \feh$_{0}$, \vsini$_{0}$) of our sample stars. We also obtain the small scale (\vt) and large scale (macroturbulence, \vmac) velocities  in the stellar atmospheres respectively, after a refinement in the minimization process to follow empirical correlations (for \vt~ we used the relation of \cite{tsantaki13} and for \vmac~ \cite{Doyle2014} for the cooler stars or \cite{valenti05} for the hotter stars). The obtained \vt~ and \vmac~ are thus not derived but calculated via relations as a function of \teff~ and \logg. In particular, fixing the macroturbulence according to the empirical relations helps us to disentangle \vmac~ from \vsini~ as they have similar convolution profiles especially for slowly rotating stars.
\item[R.1] In this step, we obtain \logg\, from spectroscopic, astrometric, and photometric data as in M22. The trigonometric gravity (\logg$_{\rm trig}$) is based on the following expression:

\begin{flalign*}\label{eq2}
\log \frac{g_{trig}}{g_{\odot}} &= \log \frac{M_{\star}/M_{\odot}}{(R_{\star}/R_{\odot})^{2}} \\ \nonumber
&= \log \frac{M_{\star}}{M_{\odot}} + 4 \log \frac{T_{\rm eff\, \star}}{T_{\rm eff \odot}} - \log \frac{L_{\star}}{L_{\odot}} 
~. ~~(1)
\end{flalign*} 
  
We derived the stellar mass (\Mstar) and radius (\Rstar) using {\em Gaia} DR3 parallaxes and photometry (including Johnson B and V synthetic band), 2MASS photometry, and the first set of spectroscopic parameters from R.0 (\teff$_{, 0}$ and \feh$_{0}$) through the isochrone fitting from the PARAM tool\footnote{\url{http://stev.oapd.inaf.it/cgi-bin/param}} \cite[][Bossini et al. in prep.]{Rodrigues2017} and including the stellar models from \cite{Moedas2022}. We repeated the spectral analysis with the \logg$_{\rm trig}$ fixed, and obtained a new set of stellar parameters (\teff$_{, 1}$, \logg$_{\rm trig}$, \feh$_{1}$, \vsini$_{1}$, \vt$_{1}$, \vmac$_{1}$). 
\item[R.2] Once the stellar parameters are obtained, we perform another iteration to re-compute the trigonometric gravity with the updated \teff$_{, 1}$ and \feh$_{1}$ values, which lead to new estimates of stellar masses and radii. We repeated the spectral analysis, keeping the new \logg~ (\logg$_{\rm trig, final}$) fixed and obtain the final set of stellar parameters (\teff$_{\rm , final}$, $\log g_{\rm trig, final}$, \feh$_{\rm final}$, \vsini$_{\rm final}$, \vt$_{\rm final}$, \vmac$_{\rm, final}$).
\end{itemize}

The convergence speed of the code depends on the proximity of the star's parameters to the initial input values at the start of the optimization process. For this dataset, the initial values are set to solar. On average, each star requires approximately four minutes per run to converge. The differences in the stellar parameters between the three different runs are illustrated in Fig.~\ref{comparison_runs} described by their median and Median Absolute Deviation (MAD). The MAD of the differences between the final two runs (Run 1 - Run 2) is lower than the average errors for all parameters in the three panels, and thus, we consider that it is sufficient to stop at the second run. Surface gravity is usually underestimated in the pure spectroscopic analysis and in particular for the K-type stars in this sample (blue points in the middle panel of Fig.~\ref{comparison_runs}). This affects the determinations of \teff~ but does not have a big impact on \feh. The difficulty to determine \logg~ from purely spectroscopic methods in particular for cool stars, is demonstrated for both methods based on the EW and spectral synthesis in previous studies \citep{Torres2012, Tsantaki2019}. However, the approach adopted in this work appears to effectively resolve this issue.

In Fig.~\ref{spectra}, we show some example cases of six stars, which were analysed with the spectral synthesis technique having a variety of rotational velocities. This plot showcases visually how spectral lines become blended, shallower and close to the continuum as the \vsini~ increases. The measurement of the EW becomes more difficult and the spectral synthesis is necessary for the analysis of this type of stars.

\subsection{Consistency between the two spectroscopic methods}

In this Section, we present our consistency checks between the two analysis methods described above. We re-computed the stellar parameters of the 187 stars in M22 using the methodology described in Sec.~\ref{sec:fasma}. The comparison between our final results and those of M22 is shown in Fig.~\ref{comparison_3rdrun}. In the three panels of Fig.~\ref{comparison_3rdrun}, we present the differences ($\Delta$) between \teff, \logg, and \feh, respectively derived in the present work and those in M22, as a function of the M22 parameters. The median difference in \teff~ is 41\,K while its MAD is 73\,K. There are four stars that show higher than 3$\sigma$ differences in \teff~ (HAT-P-14, HAT-P-41, HD209458, and WASP-94A) which are mostly located at the hotter end of the \teff~ distribution (\teff$>$6000\,K) and correspond to the orange points in Fig.~\ref{comparison_3rdrun}. In particular, HAT-P-41 shows high \vsini\, to be analysed with the EW method (22.1\,\kms) and for WASP-94A we used a higher quality SALT spectrum in terms of S/N. The agreement of the \logg~ is excellent, with a median of 0.00\,dex, and MAD=0.03~dex. We recall that in both methods the surface gravity is fixed using astrometry, photometry and the results of the spectroscopic analysis. Finally, the comparison between the two determinations in \feh~ shows a good agreement, with a median of --0.04~dex and MAD=0.05~dex. Only two stars show large discrepancies, above 3$\sigma$, in metallicity (HAT-P-14 and HAT-P-20) with the former being quite hot and the latter appears 0.34\,dex more metallic in our analysis. For this star, the average \feh~ obtained from a compilation of high-resolution studies in the literature \cite[][and references therein]{Soubiran2022} 0.21\,dex which is in-between the value of the spectral synthesis and M22 and slightly closer to the M22 value (0.10\,dex).

In Fig.~\ref{param_dependence}, we investigate possible trends that may affect our parameters with spectral synthesis as a function of \teff, \logg, and \feh. There is an overestimation in \teff\, for the 10 most metal-poor stars in our sample ($<$--0.25\,dex). The metal-poor stars show an offset of 145\,K and it is not easy to examine which of the two methods performs better at this parameter space. On the other hand, the metal-poor stars of our sample have excellent agreement in surface gravity and in metallicity (0.05\,dex 0.01\,dex, respectively). We do not notice other significant dependencies for the rest of the parameters. 

\subsection{Comparison with external samples}

We use open clusters as an external reference sample to validate our spectral synthesis method, particularly for stars that lie outside the parameter space of M22, such as hotter, fast-rotating stars. Open clusters have, as a first approach, chemical homogeneity and thus their metallicities can be well defined. Additionally, their ages can be constrained via isochrone fitting, allowing us to assess how well our results correspond to the respective isochrones. We select 22 main-sequence stars in 10 open clusters from the {\it Gaia}-ESO survey \citep[][ hereafter GES]{Randich2022}, observed with UVES (R$\sim$47\,000) in the spectral range 480-680~nm. We used the membership probability (P) based on the RV from the GES \citep{Jackson2022}, along with parallaxes and proper motions from Gaia DR3, ensuring that only stars with P$>$0.9 were included. The selected stars have rotational velocities have higher than 10\,km~s$^{-1}$, allowing for a comparison with our stars located outside the limits of M22.

For this purpose, we analysed the GES spectra with the same methodology as in Sect.~\ref{sec:fasma}, following our iterative approach to obtain the trigonometric \logg. We obtain stellar evolutionary isochrones from the PAdova and TRieste Stellar Evolution Code \citep[PARSEC version 2.1;][]{bressan12} using the clusters' metallicities and ages from the GES \citep{Jackson2022, Randich2022}. The results of our analysis are shown in the Kiel diagram in Fig.~\ref{cluster} where we present our stellar parameters and the GES results for the 22 cluster members. The stars fall very close to the isochrones of the clusters, highlighting the agreement with the theoretical predictions. Apart from the star in NGC2244, our stellar parameters are closer to the isochrones compared to the GES parameters. Overall, the median difference in \teff~ is --16\,K (MAD=82\,K) and for \logg~ is --0.07\,dex (MAD=0.12\,dex). Furthermore, we compare our metallicity determinations with the iron metallicities of each cluster provided by GES in \cite{Randich2022}. We find a very good agreement with a difference of --0.04\,dex (MAD=0.07\,dex).

\subsection{The effect of different instrument resolution}

A key objective of this work is to ensure homogeneity of the stellar parameters across the Ariel MCS. In the previous sections, we have shown the agreement of the two spectral analysis methods and the validation with open clusters. However, apart from the methods themselves, other factors could introduce inhomogeneities in the data, such as the use of different spectrographs. Ideally, the use of the same instrument with the same configuration should be adopted to maintain uniformity in the spectroscopic analysis. In practise, this is not feasible as our targets are spread in different hemispheres and to minimize the effects of resolution, we focus only on high-resolution spectrographs in the optical. Most of the spectra ($\sim$60\%) in the Ariel MCS so far come from HARPS-like spectrographs with resolution $\sim$115\,000 while the lowest resolution in our sample comes from FEROS (R$\sim$48\,000) for 12\% of our PHS. 

To investigate any discrepancies in our parameters coming from different instrument resolutions, we convolve the HARPS spectra of the M22 sample for 102 stars to the resolution of FEROS and re-derive the stellar parameters with spectral synthesis technique (Run 0 from Sect.~\ref{sec:fasma}). The comparison of the stellar parameters from the HARPS spectra with the convolved spectra (both results from Run 0), in Fig.~\ref{comparison_resolution}, shows excellent agreement: $\Delta$\teff=--7\,K (MAD=11\,K), $\Delta$\logg=0.01\,dex (MAD=0.01\,dex), and $\Delta$\feh=0.01\,dex (MAD=0.02\,dex). We can deduce that the effect of using different high-resolution spectrographs is quite small for the derivation of stellar parameters \citep[see also][]{Blanco-Cuaresma2014, Sousa2021}. 

\section{Results of the spectroscopic analysis}\label{sec:results}

\begin{figure}
\centering
 \includegraphics[width=1.\linewidth]{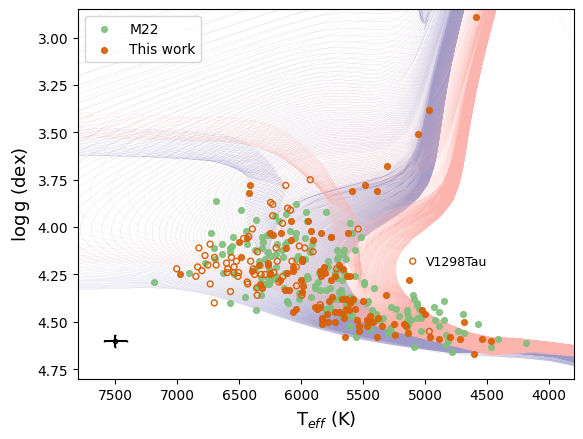}     
 \caption{Kiel diagram of our stellar sample. The stars analysed with the spectral synthesis are plotted in orange. The fast rotators of our sample are plotted with open orange symbols and the sample of M22 is plotted in green. The two grids represent the PARSEC isochrones with ages from 0.1 to 14\,Ga, with steps of 0.05\,Ga, at solar metallicity (Z\,=\,0.013, in purple) and at super-solar metallicity (Z\,=\,0.06, in red).}
 \label{hr}
\end{figure}

We present the results of our spectral analysis which consists of the new sample to be incremented to the sample of M22. We select stars among: {\em i)} the spectra considered in M22 but which could not be analysed with the EW method (including some cool slow rotators), {\em ii)} archival spectra for stars in the Ariel MCS; {\em iii)} new observations (see Sect.~\ref{sec:sample}). The new sample of this work consists of 169 FGK-type stars of which 108 are slow rotating FGK-type stars which are analysed with the EW method and 61 stars are analysed with spectral synthesis, most of the latter are F-type stars and two of them are pre-main sequence stars (see Sect.\ref{sec:special}). The consistency checks on our spectral analysis methods shown in the previous Section ensure that there are no significant trends or offsets between the two analysis methods. We can therefore, consider the results of both methods quite consistent with each other. This is crucial for the preparation for the Ariel mission, which aims to obtain homogeneous parameters for a wide range of stellar parameters. The Ariel MCS includes spectral types beyond the range where our methodologies for FGK-type stars are applicable. To address this, the Ariel ``Stellar Characterisation'' working group includes teams dedicated to the analysis of the M-type and A-type stars to be described in future works (Maldonado et al., in prep.; Ramler et al. in prep., Danielski et al., in prep.). Although the analysis methods differ for these stars, we plan to use comparison samples with overlapping stars at the cooler (late K) and hotter (early F) temperature ranges to ensure consistency. 

\begin{figure*}
  \centering
  \includegraphics[width=0.45\linewidth]{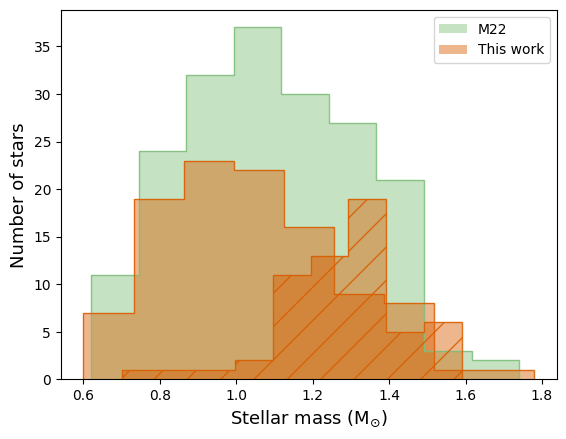}   
  \includegraphics[width=0.45\linewidth]{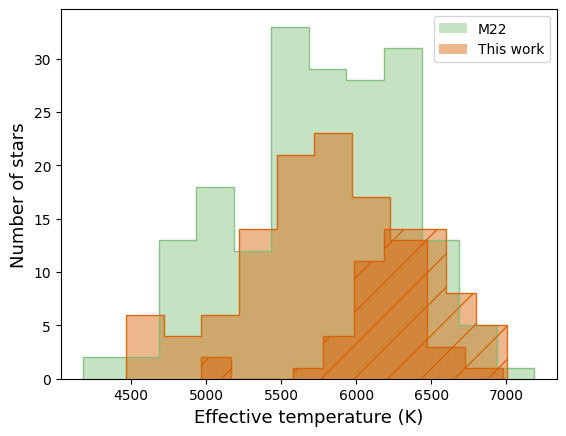} \\
  \includegraphics[width=0.45\linewidth]{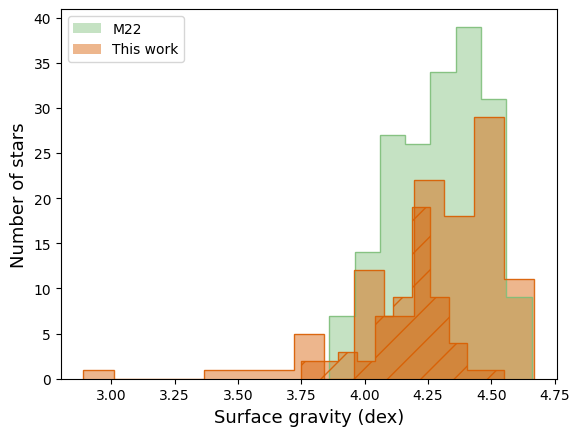}  
  \includegraphics[width=0.45\linewidth]{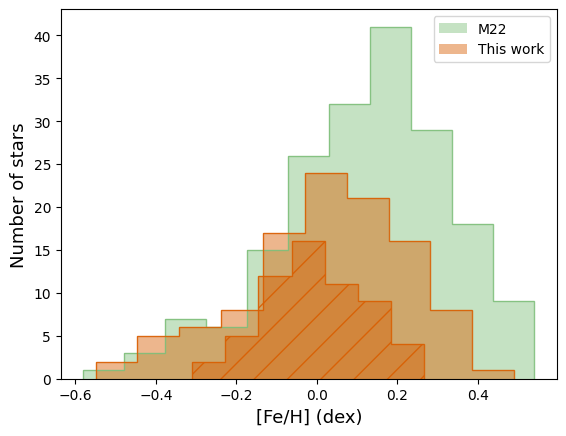}  \\
  \caption{Histograms of the distributions of the stellar parameters (Mass, \teff, \logg, \feh) for the FGK-type derived in this work (orange) and our previous work in M22 (green). The distributions of fast rotators of the present sample are shown with orange oblique stripes.}
  \label{distribution_parameters}
\end{figure*}

\subsection{Stellar parameters}\label{sec:stellar_parameters}

In the Kiel diagram of Fig.~\ref{hr}, we show the \logg\, vs \teff\, of the new samples of slow and fast rotating stars analysed in the present work and together with the sample of M22, the final sample amounts to 353 stars which represents all stars in the Ariel MCS analysed homogeneously so far. For the stars with newly obtained spectra from our observing campaigns, we provide for the first time stellar parameters in high resolution in this work. The overall sample comprehensively maps the main sequence up to \teff$\sim$7300~K, with the exception of the M-type and A-type stars for which we will devote dedicated analyses in future works. In particular, the new sample of fast rotators maps the region of the highest temperatures, and hence highest stellar masses, allowing us to extend our study of correlations between the properties of host stars and their planetary systems. We have some stars at the main sequence turn off, thus presenting the advantage of a better determination of their age (Bossini et al. in prep.). There is an excellent overlap for most stars in our sample with the grid of stellar evolutionary isochrones taken from PARSEC at solar and super-solar metallicity which is representative of the average metallicity of our sample. 

In Fig.~\ref{distribution_parameters}, we present the histograms of the distributions of the stellar parameters (\Mstar, \teff, \logg, and \feh) for our samples, together with those of the M22 sample. As already noted in Fig.~\ref{hr}, the new sample includes more massive stars, reaching up to about 1.8\,M$_{\odot}$. In contrast, the sample of FGK-type stars of this work has a peak in mass around 0.9 M$_{\odot}$, slightly lower than the peak of the M22 sample around 1.1\,M$_{\odot}$. Extending the sample to higher stellar masses is important because finding planets around these stars via RV surveys is quite difficult. Hotter stars due to their rapid rotation, greatly limit the radial velocity precision and inhibit the detection of orbiting planets \cite[e.g.][]{Galland2005}. The characterization of these systems will help us understand the observed correlations between stellar mass and planet occurrence and/or planet properties at the high mass end. The range in temperature of the two samples extends from about 4500\,K up to 7000\,K, while \logg\, spans from about 2.89\,dex to 4.67\,dex, since some sub-giant stars are included in the new version of the Ariel MCS. Finally, the new sample of FGK-type stars allows us to increase the number of stars at the low metallicity tail and the metallicity ranges between $-0.55<$\feh$<+0.49$\,dex. 

\subsection{Projected rotational velocity}

\begin{figure}
\centering
\includegraphics[width=1.0\linewidth]{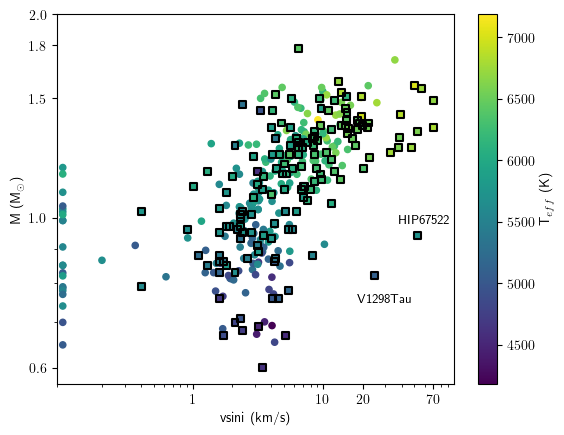}  \\ 
\caption{Stellar mass as a function of $v \sin i$ in logarithmic scales. The stellar sample presented in this work is represented with squares while circles indicate the M22 sample. The symbols are colour-coded by the effective temperature.}
\label{vsini_mass}
\end{figure}

We calculate the \vsini~ for the whole Ariel stellar sample (including M22), from the fitting of synthetic lines of specific spectral regions convolved to a rotation profile to the observed spectra as described in \cite{Gray2008} and the results are in Table~\ref{vsini_ariel}. The \vmac~ is set fixed from the empirical relations of Sec.~\ref{sec:fasma}. In Fig.~\ref{vsini_mass}, we show the relation between \vsini~ and the stellar mass, colour-coded by the effective temperature. As expected, the lower the mass, the lower the rotational velocity. In fact, the thicker radiative zone in the atmospheres of the hotter F-type stars, the higher the rotation compared to the cooler K-type dwarfs. On top of the relationship between mass and rotational velocity, stellar rotation has emerged as a promising indicator of age where at a given mass, an older star rotates more slowly than a younger one \citep[e.g.,][]{Lanzafame2015}. This can explain the presence of some low-mass stars with high projected rotation velocity, which are young (e.g. V1298\,Tau and HIP67522). 

We note that the low limit for the \vsini~ determinations with FASMA is set to zero which is a non-physical value and is related to the degeneracy between \vmac~ and \vsini~ for slow rotators. In this case, an overestimation of \vmac~ could lead to an underestimation of \vsini. Moreover, the resolution of the instrument plays a role in the precise determination of such very low velocities. 

\subsection{Kinematic properties} \label{sec:orbits}

\begin{figure*}
  \centering
  \includegraphics[width=0.4\linewidth]{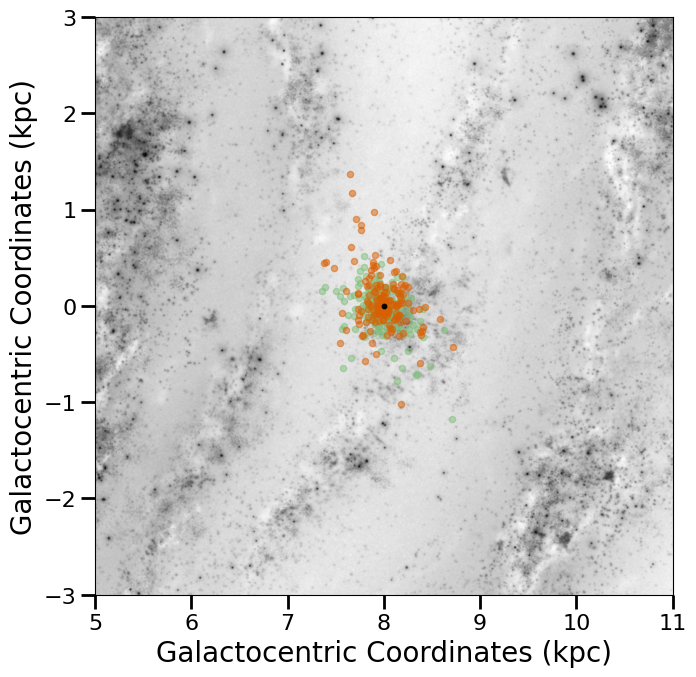}      
  \includegraphics[width=0.42\linewidth]{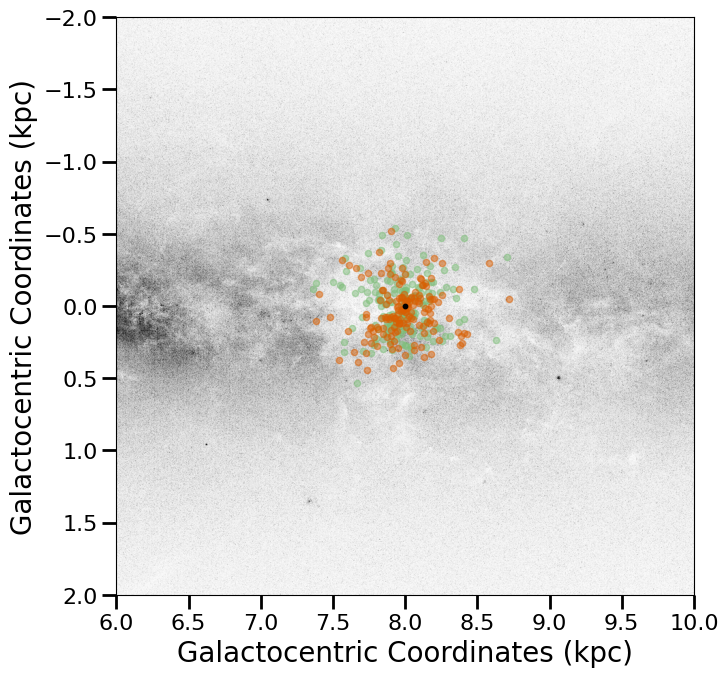}   
  \caption{Galactic positions (left panel: face on and right panel: edge on) of the Ariel MCS analysed in this work for the two sub-samples: this work (orange points) and M22 (green points). The face on background image of the Galaxy is taken from NASA/JPL-Caltech (R. Hurt SSC/Caltech) and the face off from ESA/Gaia/DPAC.}
  \label{galactic_coo}
\end{figure*}

\begin{figure}
  \centering
  \includegraphics[width=1\linewidth]{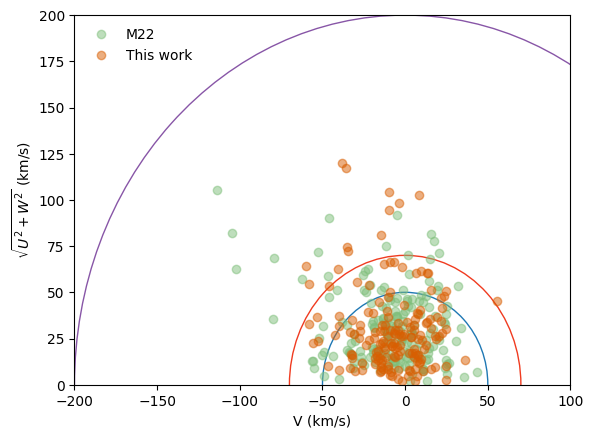}   
  \caption{The separation of our sample in the Galactic disc based on their kinematics. Stars within the blue circle belong to the thin disc, stars within the red annulus are transitioning between thin and thick, and stars within the purple belong to the thick disc. Symbols are colour-coded as described in Fig.~\ref{galactic_coo}.}
  \label{kinematics}
\end{figure}

The Ariel targets in our sample are located in the solar neighbourhood, as shown in Fig.~\ref{galactic_coo}\footnote{The images were generated by the python package, mw-plot: \url{https://milkyway-plot.readthedocs.io/en/stable/index.html}} in which our samples are projected on (left panel) and in front (right panel) the Galactic plane. From Fig.~\ref{galactic_coo}, we can discern that Ariel targets are mainly located near the Local spiral arm, and that they are at low altitudes on the Galactic plane.  

Stars belonging to the solar neighbourhood can be further grouped in distinct stellar populations, characterised by different kinematic and chemical properties: the thin and thick discs and halo. In this work, we use kinematic criteria to classify the stars of our samples in the above populations. In fact, the velocity components in Galactic coordinates (U, V, W) and the orbital parameters allow the stars to be separated into the main Galactic constituents. We computed the velocity components and the orbital parameters with the {\sc galpy} package \citep{Bovy2015}\footnote{\url{https://www.galpy.org/}}, using as input the {\em Gaia} DR3 data (positions, parallaxes, radial velocities, and proper motions). We assumed, as in M22, the model {\sc MWpotential2014} for the gravitational potential of the Milky Way. The local standard of rest (LSR) velocity was set to V$_{\rm LSR}$=220\,\kms \citep{bovy12}, and we assumed (U, V, W)$_{\odot}$~=~(11.1, 12.24, 7.25)~\kms~ for the velocity of the Sun relative to the LSR \citep{Schonrich10}. 

In Fig.~\ref{kinematics}, we show the distribution of our targets in the Toomre diagram, separating them at a first approximation between thin and thick disc stars using the same kinematic criteria as in M22. Most of our sample stars belong to the thin disc (|V$_{\rm tot}$|\footnote{The total velocity is defined as: V$_{\rm tot}$=(U$^{2}$+V$^{2}$+W$^{2}$)$^{1/2}$}~$<$~50~\kms), a small sample is in the transition region between thin and thick disc (50~\kms~$<$~|V$_{\rm tot}$|~$<$~70~\kms) and there are few stars part of the thick disc (70~\kms~$<$~|V$_{\rm tot}$|~$<$~200~\kms). There are no stars that belong to the halo (|V$_{\rm tot}$|$>$~200~\kms). The orbital properties of our sample are presented in Table~\ref{orbital}.

\subsection{Special cases}\label{sec:special}

In our sample, we have two low-mass, pre-main sequence stars (HIP67522 and V1298Tau). These are very young systems with high rotational velocities. The host star HIP67522 is an early G-type member of the Sco-Cen association. V1298 Tau, on the other hand, is a young K-type star belonging to the Group 29 stellar association \citep{David2019}. Both stars do not show emission in their spectra even though their strong surface magnetic fields can affect their line profiles and therefore, their stellar parameters. Our spectral analysis is consistent with literature values, indicating that the spectral lines we used are not greatly affected by these phenomena \citep{Finociety2023, Turrini2023b}. 

The fastest rotators in our sample are KELT-7 and KOI-12, both exhibiting rotational velocities of 70.5\,\kms. The robustness of FASMA's spectral analysis has been tested for \vsini\, values around 50\,\kms. After a visual inspection on the fits of the synthetic spectra with the observations, we confirm the validity of the parameters. Furthermore, we compare our parameters (KELT-7: \teff=6685$\pm$90\,K, \logg=4.20$\pm$0.02\,dex, \feh=0.13$\pm$0.05\,dex and KOI-12: \teff=6800$\pm$135\,K, \logg=4.23$\pm$0.04\,dex, \feh=--0.10$\pm$0.07) with literature values obtained from high-resolution analyses. Specifically, we find good agreement within uncertainties when comparing our parameters with those from \cite{Bieryla2015} for  KELT-7 (\teff=6789$^{+50}_{-49}$\,K, \logg=4.149$\pm$0.019\,dex, \feh=0.139$^{+0.075}_{-0.018}$\,dex) and  from \cite{Bourrier2015}  for KOI-12 (\teff=6820$\pm$120\,K, \logg=4.25$\pm$0.15\,dex, \feh= 0.09$\pm$0.015\,dex). Given this consistency, we include these stars in our analysis. 

We could not provide mass determination for HAT-P-13 because the convergence of the mass was not achieved from the isochrone fitting and thus, we present only the spectroscopic parameters for this star in Table~\ref{sample_params}. Moreover, we provide an update on two stars on the parameters of M22 which were re-observed with SALT providing higher quality spectra (WASP-94A and WASP-79). Finally, from the comparison of the spectral synthesis results with the parameters of M22, HAT-P-14 appears as an outlier in both \teff\, and \feh. As discussed in Sect.~\ref{sec:method}, the literature values are in better agreement with the parameters from spectral synthesis which indicate that the values in M22 were inaccurate. We update also the parameters for HAT-P-14 in Table~\ref{sample_params} as well. 
 
\section{Relationships between stellar and planetary properties}\label{sec:planets}

\begin{figure}
  \centering
  \includegraphics[width=1.0\linewidth]{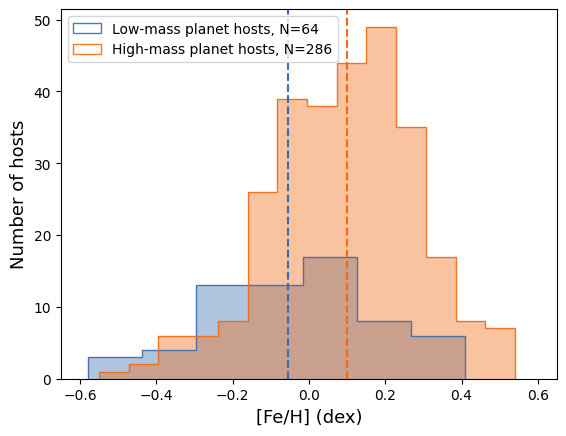}  
  \caption{Metallicity distributions of PHS with low-mass (blue) and high-mass planets (orange). The vertical lines represent their median metallicity values.}
  \label{fig:dist_metal_plmass}
\end{figure}

\begin{figure}
  \centering
  \includegraphics[width=1.0\linewidth]{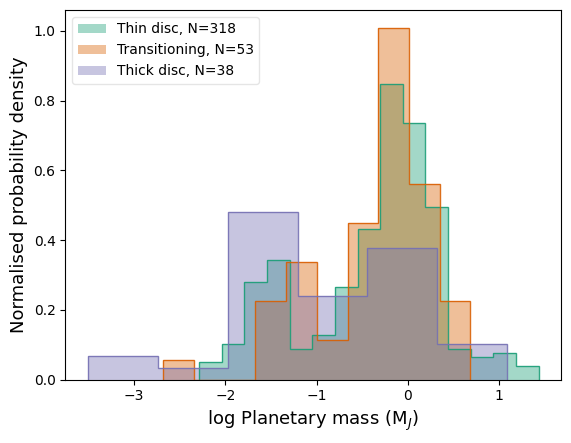}  \\
  \includegraphics[width=1.0\linewidth]{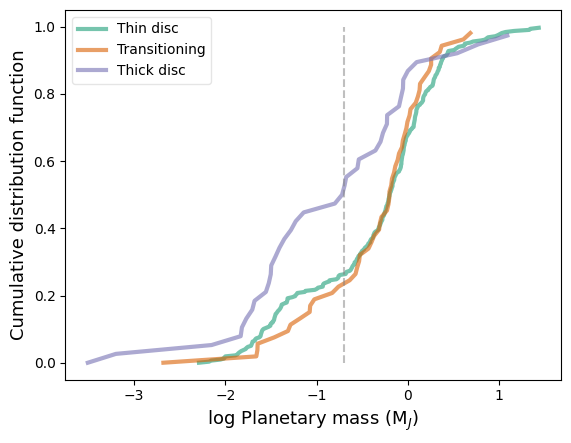}  
  \caption{Upper panel: Distribution of planetary masses separated into the three stellar populations: thin, thick disc and thin/thick transitioning. Bottom panel: The cumulative distribution function of the three planetary mass populations. The gray dashed line shows the limit of 0.2\,\Mj~ to separate between high and low mass planets.}
  \label{fig:dist_planetmass_discs}
\end{figure}

The Ariel MCS so far contains fully homogeneous stellar parameters for 353 stars combining this work and M22\footnote{As mentioned in Sect.~\ref{sec:special}, we obtained new spectra for two stars which were already analysed in M22 and updated the parameters for one star in M22.}. After a cross match with the NASA Exoplanet Archive\footnote{\url{https://exoplanetarchive.ipac.caltech.edu}}, we find 446 planets orbiting our hosts. Currently, 368 of these planets are in the Ariel MCS. In this Section, we study the connection between stellar and planetary properties, more specifically the impact of stellar mass and stellar \feh\, on planetary mass (\Mp), radius (\Rp), orbital properties, and planet multiplicity for planetary systems. While multiplicity was retrieved from the NASA Exoplanet Archive, the planetary radius and mass were retrieved from the Ariel MCS. We refer to \cite{Edwards2019,Edwards2022} for more details on how the list was compiled from the NASA exoplanet catalogue. Finally, we note that the planetary properties used here, being a collection from the literature, are not determined in a uniform way. The parameter space of the planetary properties is shown in Fig.~\ref{planet_param_fig}. 

\subsection{The effect of stellar metallicity on exoplanet populations}

From the first discoveries of exoplanets, the positive correlation of giant planet frequency with stellar iron metallicity has been established \citep[see review by][]{Adibekyan2019}. To investigate the effect of stellar metallicity on planet populations, we divide our sample into stars that host high- and low-mass planets as in M22. We note that there is no unambiguously accepted low-mass limit to distinguish high-mass planets from low-mass ones. Different values are proposed in the literature and usually range between 0.09-0.3\,\Mj~ \cite[e.g.,][]{Howard2012, Hatzes2015}. In the present work, we use the value 0.2\,\Mj~ to separate the two groups, which represents well the planet-mass gap \citep{Ida2004} in the left panel of Fig.~\ref{fig:mass_radius_multi} (see next Section) and, moreover, it is close to the average of the literature values. In Fig.~\ref{fig:dist_metal_plmass}, we show the distribution of the stellar metallicity of the hosts for the two sub-samples. In case of multiple systems, we consider only the most massive planet. The average metallicities of the two different groups are very different: --0.06\,$\pm$0.03\,dex for the stars hosting only low-mass planets and 0.10\,$\pm$\,0.01\,dex for the hosts with  at least one high-mass planet. The two-sample Kolmogorov–Smirnov (KS) test shows that there is a tiny probability that the two sub-samples have the same underlying metallicity distribution (KS statistic\,=\,0.283 with p-value\,=\,3.5$\times$10$^{-4}$). This result comes as a confirmation to the extensive observational and theoretical works in the literature to support that giant planets orbit more metallic stars than lower mass planets. The average metallicity value in the solar neighbourhood for FGK-type single dwarfs is --0.115\,$\pm$\,0.003 dex \citep[e.g.,][]{Osborn2020}, which is in agreement with the metallicity of our hosts with exclusively low-mass planets. This suggests that they do not exhibit a strong preference for iron metallicity, at least for systems in the solar vicinity.

The stars in the solar neighbourhood are part of the Galactic disc which is separated into two main components: the thin and thick discs, characterised by stellar populations with different ages and chemical composition, reflecting different star formation histories \citep[e.g.,][]{Bensby2005, Adibekyan2012}. In principle, the varying formation conditions of host stars in different populations may have also contributed to differences in their planetary systems. In Sect.~\ref{sec:orbits}, we showed the distribution of our sample across the two discs, with a large prevalence of stars in the thin disc, and a lower number of stars in the thick disc. The mean metallicity of the thin disc stars in our sample is 0.09\,$\pm$\,0.01\,dex and of the 29 thick disc stars is --0.02\,$\pm$\,0.05\,dex. This reflects the Galactic chemical evolution with the PHS in the thick disc being less metallic than the hosts in the thin disc. However, both values are significantly different from the average values of field stars as reported in \cite{Adibekyan2012} with --0.06\,dex and --0.60\,dex, respectively. This difference likely reflects the tendency for stars with higher metallicities to host planets, while stars with low metallicities show a lack of planets. Observationally, only a few planets have been confirmed orbiting stars with \feh\,$<$\,--0.60\,dex in the literature.

Given that the two components of the disc have distinct metallicities— with the thick disc being more iron-poor and $\alpha$-enhanced— we expect their planets to also reflect these chemical differences in their properties. In Fig.~\ref{fig:dist_planetmass_discs}, we investigate the differences of planetary systems in terms of their masses in the thin and thick discs, even though our sample is statistically incomplete. Lower-mass planets in our sample are found orbiting thick disc stars which are more iron-poor while high-mass planets are more abundant in the younger more metal-rich thin disc. We performed a two-sample KS test based on the cumulative distributions of Fig.~\ref{fig:dist_planetmass_discs} which suggests significantly different planet mass distributions for the thin and thick disc samples with KS statistics of 0.311 and p-value\,=\,2.0$\times$10$^{-3}$. While the thin disc and the thin/thick disc transition region in our sample are dominated by high-mass planets (74\% and 75\%, respectively), the thick disc low-mass planets comprise less than half of them (47\%). Our results show that in our sample more massive planets are formed predominantly around stars with solar or super-solar \feh~ and are mainly located in the thin disc. Generally, thin-disc stars are younger than thick-disc stars, and due to the relationship between age and metallicity \citep[e.g.,][]{Sahlholdt2022MNRAS.510.4669S}, those richest in metals should also be the youngest. Therefore, there could also be a relationship between the presence of giant planets and stellar age: they might be likely found orbiting younger stars \citep[e.g.,][]{Swastik2022}. 

Recent works have used simulations of planet formation in the thin and thick disc confirming variations on planet populations in different regions of the Galactic disc \citep{Nielsen2023, Boettner2024}. In particular, giant planets are most common around thin disc stars since these stars have an overall higher budget of solid particles. Giant planets are very rare (less than 1\%) around thick disc stars. Low-mass planets, on the other hand, are expected to be found in all parts of the Galaxy because their formation depends less on stellar metallicity. In addition, the same studies show that the average age of the planet population around Sun-like stars in the thin disc is older than the thick disc and correlates with planetary mass with higher planet masses being younger which agrees with our findings. However, based on observations alone, it is challenging to disentangle the effects of different star formation histories from biases introduced by detection methods when examining the relationship between planetary and stellar properties. Moreover, in order to define occurrence rates in the Galactic disc, there is a need for volume-limited samples which is not our case. Nevertheless, the preference of low-mass planets in the thick disc is a consequence of the iron metallicity correlation with the planet presence and is also evident in this work.

Observational works have shown that there are differences in the rate of planet formation between thin and thick disc stars \cite[e.g.,][]{Haywood2009ApJ...698L...1H, Adibekyan2012a, Bashi2022}. For instance, \cite{Bashi2022} found that thick-disc stars have a lower occurrence rate of close-in super-Earths compared to metal-rich thin-disc stars, suggesting that the masses of planets could be different between the different stellar populations based on the HARPS GTO planet search programme. But in a low-iron regime, small-sized planets are mostly enhanced by $\alpha$ metals and belong to the Galactic thick disk \cite{Adibekyan2012b}. Recently, \cite{Biazzo2022} also confirmed that stars hosting low-mass planets tend to belong to a thicker disk using a smaller but well-characterised sample under the GAPS programme at the TNG. The above works used both chemical and kinematic data to separate the Galactic components. A separation based also on chemistry (namely based on their [$\alpha$/Fe] abundances) is more reliable, because chemistry is a relatively more stable stellar property throughout the life of a star than the spatial positions and kinematics. Along with the CNO determinations already presented in \cite{daSilva2024}, in our future work, we will provide abundances for other $\alpha$-elements as well for the Ariel MCS (Delgado-Mena in prep.) and a more reliable picture on the planet distributions in the Galactic disc. However, even with kinematics alone, we have shown here that Galactic chemical evolution, along with stellar properties such as mass and metallicity, can play a significant role in shaping the resulting planetary population.

\subsection{Relationship between stellar mass and planetary mass and radius}

\begin{figure*}
  \centering
  \includegraphics[width=0.48\linewidth]{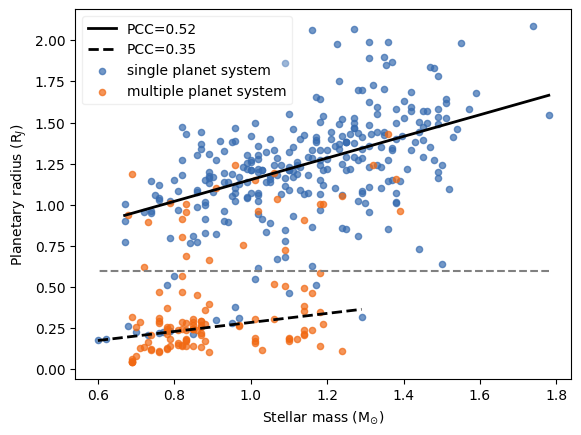} 
  \includegraphics[width=0.48\linewidth]{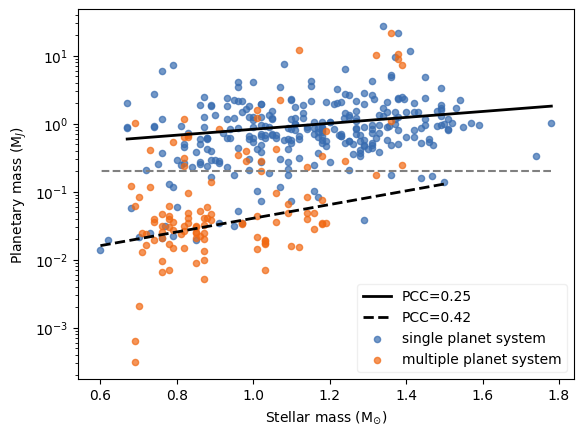} 
  \caption{Left panel: Planet radius as a function of stellar mass. The orange points indicate planets in multiple systems and the blue points are single planet systems. The dashed lines separates the smaller planets at 0.6\,\Rj from giant planets. Right panel: Planetary mass (in logarithmic scale) as function of stellar mass. The grey dashed lines separates the low-mass planets at 0.2\,\Mj from giant planets. The black lines represent the linear fits with their Pearson Correlation Coefficients between planetary mass and radius with stellar mass for giant planets (solid black line) and low-mass planets (dashed black line).}
  \label{fig:mass_radius_multi}
\end{figure*}

In Fig.~\ref{fig:mass_radius_multi}, we show in the left panel the relationship between the planetary radius and the stellar mass, while in the right panel the relationship between the planetary mass (in logarithmic scale) and the stellar mass. In the figure, we highlight the single-planet systems and the ones having more than one planet. In both panels, the gray dashed horizontal lines separate the massive and larger planets from the low-mass and smaller ones with 0.6\,\Rj~ and 0.2\,\Mj~ set as limiting values, respectively. We investigate the correlations between the stellar mass and the properties of the planet with linear fits. We used their Pearson correlation coefficients (PCC) as a diagnostic for the significance of the relationships. To compute the linear fits, we do not separate planets in those belonging to single or multiple systems. 

The correlation is stronger between the \Rp~ and stellar mass for the larger-sized planets (PCC=0.52 with p-value=3.7\,$\times$10$^{-22}$) than for the smaller-sized ones (PCC=0.35 with p-value=1.5\,$\times$10$^{-4}$). \cite{Lozovsky2021} suggested that this correlation, at least for the smaller-sized planets might be the result of different planetary composition and/or structure of the planets surrounding different stellar types. Indeed, the authors found a clear trend where planets around more massive stars have higher H-He mass fractions which impact the planetary radii. Their sample was comprised of GK-type stars but we also confirmed this correlation exceeding to FGK-types both in M22 and in this work with the extended sample. We note, however, that we do not have small-sized planets for \Mstar$>$ 1.3\,\Msun~ hence, this correlation applies until late F types.

The masses of low-mass planets show moderate correlation with stellar mass while for high-mass planets it is weaker (PCC=0.42 and PCC=0.25, respectively with their p-values equal to 2.8\,$\times$10$^{-6}$ and 1.7\,$\times$10$^{-5}$). The correlation of \Rp~ with stellar mass seems to be more evident for both large and small sized planets and while \Mp~ is moderate only for low-mass planets and weak for high-mass planets. Theoretical works of synthetic planet populations around low-mass stars (0.1-1.0\,\Msun) confirm that planetary masses increase with the host stellar masses \citep{Burn2021}. However, the planetary mass does not linearly scale with the stellar mass despite the linear scaling of the gas and solid disk mass but possibly scales as a power law \citep{Alibert2011} which could be the reason for the low PCC for \Mp. 

Another interesting aspect in Fig.~\ref{fig:mass_radius_multi} is the absence of low-mass and small planets around stars with masses higher than 1.3-1.5\,\Msun. This could also be partially attributed to an observational bias, as it is more challenging to detect small-sized planets around hotter and brighter stars. Their transits are shallower, making them harder to identify around hotter stars than around cooler ones. Also, the RV detection method has its limitations on the discovery of low-mass planets on the more massive stars because of their fast rotation. In fact, according to planet population synthesis models, the occurrence rates in Super-Earth, Neptunian, and giant planets increase with increasing stellar mass \citep[e.g.,][]{Boettner2024}. 

\subsection{Relationship between stellar mass and planet multiplicity}

\begin{figure*}
  \centering
  \includegraphics[width=0.48\linewidth]{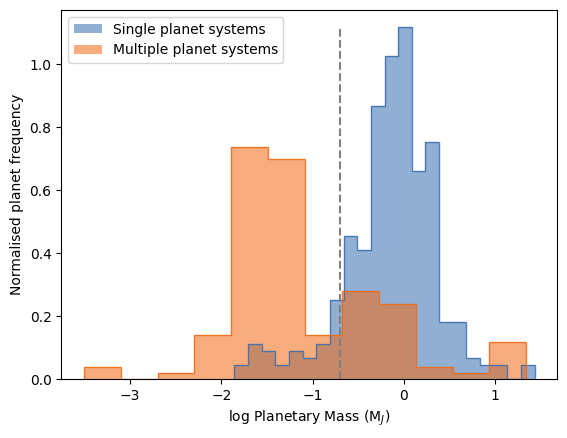} 
  \includegraphics[width=0.48\linewidth]{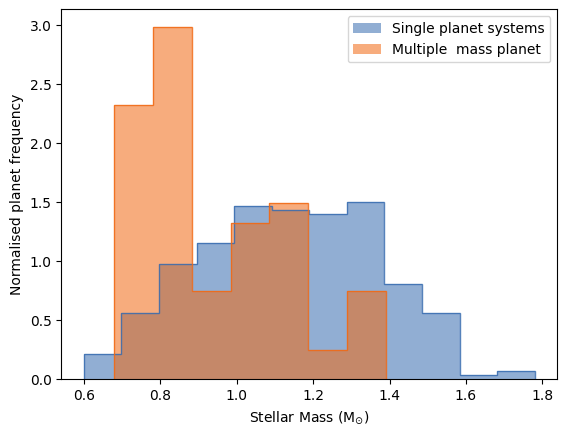} 
  \caption{Left panel: The planet mass distribution of the Ariel MCS analysed in this work divided into planets belonging in single (blue) and multiple (orange) systems. The vertical horizontal line separates the high and low-mass planets at the limit value of 0.2\,\Mj. Right panel: The stellar mass distribution of the Ariel MCS analysed in this work divided into planets belonging in single (blue) and multiple (orange) systems.}
  \label{fig:multi}
\end{figure*}

Figure~\ref{fig:mass_radius_multi} also provides information on the relationship with the multiplicity of the planetary system and the mass of the host star. In terms of multiplicity, we find 151 exoplanets, 119 of which have planetary mass determinations, belonging to a total of 55 different multiple-planet systems. The majority of multi-planet systems in our sample are populated by low-mass planets, comprising 76\% (42/55 systems). Systems with at least one high-mass planet amount to 42\% (23/55 systems) in the sample (see left panel of Fig.~\ref{fig:multi}), while the remaining 58\% (32/55) consist solely of low-mass planets. In contrast, 89\% of single systems (263/295) host a high-mass planet which is also close-in (semi major axis of the orbit $<$\,1\,au). Most low-mass planets in our sample belong to multiple systems orbiting lower-mass stars (see right panel of Fig.~\ref{fig:multi}). Since most of the low-mass planets orbit low-mass stars and low-mass planets are usually located in multiple systems, we can deduce that in our sample planet multiplicity decreases with increasing stellar mass, i.e. spectral type \citep[see also][]{Yang2020}. The tendency for close-in giant planets to be less common around low-mass stars \citep{Johnson2010, Mulders2015} and the tendency for small planets to be more common around cooler stars \citep{Howard2012} have been investigated in surveys dedicated to planet detection with both the transit and RV methods and could explain the bi-modality of the histograms in the left panel of Fig.~\ref{fig:multi}. 

Detection biases in both RV and transit surveys generally make it more challenging to detect lower-mass and smaller planets. As a result, a number of single-planet systems is expected to be misclassified because of undetected companions. The population of single systems in our sample is composed mostly of close-in giants. Theoretical works on planetary population synthesis models show that the expected multiplicity for massive close-in planets is 1 \citep[one planet per system,][]{Emsenhuber2021}. On the other hand, \citet{Zhu2022} developed a method to recover the intrinsic multiplicity distribution of planets from observational surveys and also found the average intrinsic multiplicity (the average number of planets per planetary system) of such planets to be almost consistent with unity (1.08 $\pm$ 0.07). Other observational studies confirm as well that planet multiplicity tends to decrease for systems that host more massive planets \citep{Latham2011}. Hot Jupiters in particular rarely have nearby companions \citep{Steffen2012}, though about half are accompanied by more distant, high-mass planets \citep{Knutson2014}. The above works indicate that the number of undetected smaller planets in this sample comprised of close-in giants should be small. However, a comprehensive understanding of planet multiplicity requires comparing single and multiple systems while accounting for stellar parameters (e.g., \teff, \feh, and \Mstar) and the probability of planet detection.

This demography characterising our sample is probably related to different evolutionary paths of the systems hosting only low-mass planets and the systems enclosing giant planets too. The history of planetary systems is, indeed, shaped by their ``primordial evolution'', which accounts for the dynamical interactions when young planets are still embedded in their native circumstellar disc, and their ``secular evolution'', which refers to the phases of the system after the disc has dissipated. During their formation and primordial evolution, planets undergo orbital migration due to disc-planet interactions, which in multi-planet systems can lead to their convergent migration and resonant capture on stable or temporary stable configurations (see the reviews by \citealt{WinnFabrycky2015, ZhuDong2021} and references therein). After the disc dispersal, the secular evolution of multi-planet systems is governed by the gravitational interactions among the planets. In those systems where the initial architectures are unstable such interactions can trigger phases of dynamical instability and planet-planet scattering (e.g. \citealt{chambers2001,laskarPetit2017}). These instabilities appear to be frequent among the known multi-planet systems \citealt{laskarPetit2017,gajdoVanko2023}) and are believed to be the cause of the observed paucity of resonant systems among exoplanets (\citealt{WinnFabrycky2015,ZhuDong2021}).

In multi-planet systems hosting giant planets these dynamical processes will lead to more destructive outcomes which will more often result in the loss of planetary bodies and diminished multiplicities \citep{ZinziTurrini2017,Turrini2020,Burn2021,Schlecker2021}. Specifically, migrating giant planets destabilise or accrete all smaller planetary bodies they encounter in their pathway (e.g., \citealt{shibata2020,Turrini2021}), leaving dynamically excited planetesimal disks where collisions are responsible for transforming planetesimals back to dust \citep{Bernabo2022,Turrini2023}. Systems hosting giant planets that do not undergo such destructive processes will still be characterised by stronger gravitational perturbations than their lower mass counterparts. These massive systems will have higher chances of experiencing dynamical instabilities, removal of planets across their secular evolution \citep{Burn2021, Schlecker2021}, and/or the scattering of planets on high inclined orbits causing them to not transit to our eye \citep{laskarPetit2017}.

As a result, planetary systems hosting giant planets are more likely to become or appear as single-planet systems over the course of their life, in agreement with what shown by our sample (Fig.~\ref{fig:mass_radius_multi} and left panel of Fig.~\ref{fig:multi}). Conversely, systems hosting only lower-mass planets will be characterised by less violent and destructive formation and evolution histories, leading more often to higher observable multiplicities. So, going back to the results shown in Fig.~\ref{fig:mass_radius_multi} and Fig.~\ref{fig:multi}, we can conclude that the relationship between the mass of the star and the mass/radius of the planet favours giant planets around massive stars, as discussed above. In such cases, the dynamical evolution of the planetary system favours the formation of a single planet system, as shown by the data and expected by the models of planet interactions \citep[see, e.g.][]{WinnFabrycky2015}. For low-mass stars, in whose planetary systems a single giant does not dominate, multiplicity is instead generally maintained. 

\subsection{The role of metallicity on planetary radius as a function of stellar mass}\label{stellarmetal_planetradius}

\begin{figure}
  \centering
  \includegraphics[width=0.95\linewidth]{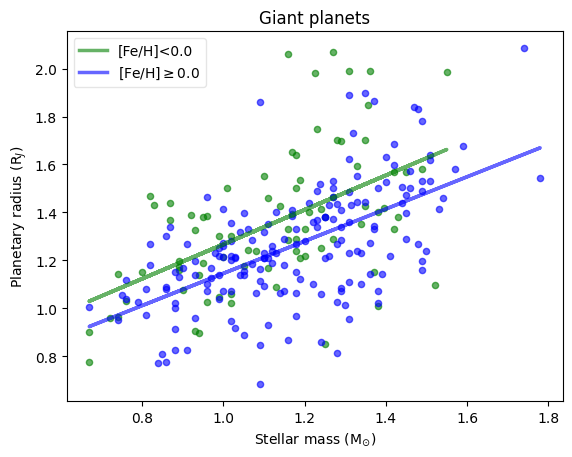}
  \caption{Planet radius of giant planets in the sample as a function of stellar mass derived in this work divided into two metallicity bins. The planets in metal-rich hosts (\feh\,$\geq$\,0.0\,dex) are shown in blue and the metal poor (\feh\,$<$0.0\,dex) in green. The linear fits to the two datasets are colour-coded in the same way.}
  \label{fig:stellarmass_planetradius_giants_2bins}
\end{figure}

As in M22, we investigate the role of metallicity in the relationship between planetary radius and stellar mass by extending the sample to larger stellar masses. We perform our analysis on the single-planet systems hosting only giant planets (\Mp\,$\geq$\,0.2\,\Mj~ and \Rp\,$\geq$\,0.6\,\Rj) which show the strongest correlation in the left panel of Fig.~\ref{fig:mass_radius_multi}. Following the same prescription as in M22, we divide the sample into three metallicity bins that represent sub-solar (N=17), solar (N=111), and super-solar values (N=132) to include the effect of stellar metallicity. In Fig.~\ref{fig:stellarmass_planetradius_giants}, we plot \Rp~ as a function of stellar mass for a total of 260 stars/systems with a single known giant planet. We observe a similar trend between \Rp~ and \Mstar~ and \feh~ as reported in M22 (their Fig. 13), indicating that, at a given \Mstar, giant planets around low-metallicity stars are more inflated than those orbiting around higher metallicity stars. However, statistical analysis using F-tests (based on the comparison of residual variances) does not reveal significant differences between all three linear fits, implying that these trends should only be interpreted qualitatively. Nevertheless, when dividing the metallicity range into two equally sized bins (\feh\,$\geq$\,0.0\,dex and \feh\,$<$\,0.0\,dex) shown in Fig.~\ref{fig:stellarmass_planetradius_giants_2bins} we find statistically significant differences (p-value = 0.01), confirming the relation between \Rp, \Mstar, and \feh, where at a given \Mstar, giant planets around low metallicity stars are more inflated than those orbiting around higher metallicity stars. We show that the correlation is still valid for our expanded sample of stars and for stars more massive than 1.5\,\Msun~ as well, which was the upper mass limit in M22. With this work, 11 additional stars with masses greater than 1.5\,\Msun~ have been added to our sample, increasing their total to 14 stars. With our new analysis, we can extend its validity to higher masses up to about 1.8\,\Msun~ while we could not find an equivalent clear relationship for \Mp~ (see Fig.~\ref{fig:stellarmass_planetmass_giants}). A possible explanation given in M22 is that, after removing the effect of stellar mass, the giant planets around more metallic stars are able to accrete larger amounts of heavy elements as they form to be denser and, therefore, with smaller radii. Similar results were also obtained from \cite{Biazzo2022} where the authors analysed uniformly a sample of transiting PHS and studied, among others, the relations between planetary mass, radius, density, and \feh. 

We also study the subset of systems with a stable evolution which is described by the eccentricity ($e$) of the planetary orbit i.e., with $e < 0.1$, that plausibly preserved the signatures of the primordial architecture (see next Section). For this subset of 232 planets with low eccentricity ($e < 0.1$) the slopes of the regression lines quantitatively change, yet the global relation remains qualitatively unaltered and the data spread is reduced (see Fig. \ref{fig:R_M_feH_elt01}). From these particular relationships we see that secular evolution of planetary systems can alter the system architectures, but such alteration does not erase the primordial correlations with the host star properties. At zero order, even when we ignore the evolution of the system, correlations are still visible. It must be stressed, though, that for a robust quantitative study one needs to use homogeneous planetary parameters where the values are accurately estimated. In conclusion, these results indicate that some relationships between planetary and stellar properties can be studied even in the presence of altered evolved architecture. Other relationships, for instance the study of planetary occurrence rates and multiplicities, require either to use systems with the unaltered architectures, or systems which had minimal dynamical alterations. 

\subsection{Setting the ab-initio planetary architecture with planetary eccentricity}

\begin{figure*}
\centering
    \includegraphics[width=1\linewidth]{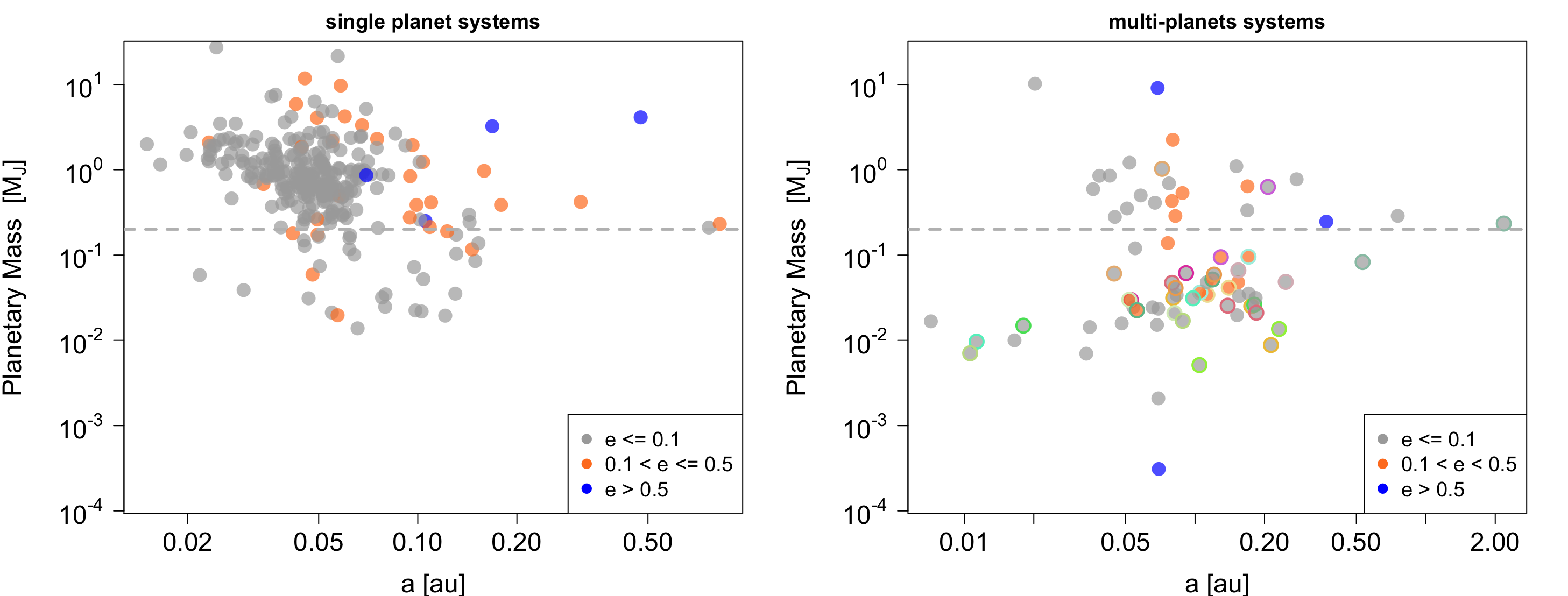}
    \caption{Planetary mass versus semi-major axis and the eccentricity of the planetary orbit. The left panel includes the single planet systems while the right the multi-planetary systems. The colour of the points in both panels corresponds to the planetary eccentricity as defined in the legends. In the right panel, the colour of the edges of the points represents the planets within the same system and the points with no different colour at the edges corresponds to planets which are the only companions in their system that made it into the Ariel MSC. The dashed line marks the adopted mass limit to define a giant planet (see text for more details).}
    \label{fig:M_a_ecc}
\end{figure*}

The dynamical evolution of the planetary systems can be investigated by studying the eccentricities of planets that impact the original architectures of the planetary systems. In Fig.~\ref{fig:M_a_ecc}, we show our planetary sample divided into single-planetary and multi-planetary systems in the parameter space of planetary mass versus semi-major axis. Marked eccentricities (e.g. $e > 0.1$) are the signpost of violent dynamical events such as instabilities or planet-planet scattering involving planetary bodies of comparable masses \citep[e.g.,][]{chambers2001, ZinziTurrini2017, laskarPetit2017, Turrini2020}. In the single-planetary sample 11\% of planets have $e > 0.1$, and among these 82\% are giant planets. On the other hand, in the multi-planetary ones, 28\% of planets have $e > 0.1$, of which 53\% are giants. The fact that the majority of planets on eccentric orbits in single-planet systems are giant planets suggests that their systems originally hosted additional giant planets. This is supported by the fact that all giant planets on eccentric orbits in our sample that belong to multi-planet systems have other giant planets as planetary companions. Conversely, giant planets in multi-planet systems where the other bodies are low-mass planets are all on circular orbits ($e = 0$). 

Multi-planet systems containing low-mass planets on eccentric orbits can host both low-mass and giant planets as planetary companions. This suggests that single-planet systems hosting low-mass planets on eccentric orbits either contain undiscovered planets, or originally possessed additional low-mass planets (as the surviving planet could not have caused the loss of a more massive one). This discussion highlights how the multiplicity and architecture of their observed systems can differ from the original native set-up. Consequently, the dynamical evolution needs to be taken into account when examining possible correlations with the host star properties as in Sect.~\ref{stellarmetal_planetradius}. 

\section{Summary and conclusions}\label{sec:summary}

In this work, we present a methodology to determine atmospheric parameters for fast rotating, mostly F-type stars which are part of the target list of the Ariel space mission. This method is based on the spectral synthesis technique and to ensure consistency with our previous work which uses the EW method, we re-analysed a sample of FGK-type stars from M22 by following as closely as possible the same prescriptions, such as model atmospheres, radiative transfer code and atomic line data. Moreover, we kept the surface gravity fixed to trigonometric values. The comparison of the two methods shows excellent agreement in all stellar parameters. We then analysed a sample of 36 rotating stars and additional 25 ones which did not converge with the EW method, using the spectral synthesis technique. Finally, we added 108 FGK stars analysed with the EWs method of M22 as a continuation of our first catalogue. With the present work, we almost doubled the sample of PHS with homogeneous stellar atmospheric parameters and stellar masses that will be observed by the Ariel satellite. 

We added the 169 stars of this work to the 187 stars from M22 and analysed the global properties of the full sample, including the distribution of the stellar parameters, their kinematics and orbits. We identified a larger population belonging to the thin disc, and a few stars likely belonging to the thick disc. We also explore the star-planet relationship and find the following: 

\begin{itemize}
    \item[-] Our sample reflects the findings that high-mass planets orbit more metal-rich stars which belong to the thin disc. The lower-mass planets can be found also in more metal-poor environments and are more likely to be hosted around thick disc stars. 
    \item[-] There is a moderate correlation between planetary mass and stellar mass and a stronger one between planetary radius and stellar mass in particular for the high-mass planets.
    \item[-] Low-mass planets which are more often found around low-mass stars also belong in multiple planetary systems.
    \item[-] We confirm the correlation between the stellar mass and the radius of the planet, with larger planets orbiting around more massive stars and larger planets around more metal-poor stars at a given stellar mass, found in M22 and expanding here to higher stellar masses.
    \item[-] Multiplicity and architecture of planetary systems as inferred from their eccentricities can differ from the original native set-up and the dynamical evolution needs to be taken into account when examining possible correlations with the properties of PHS. 
\end{itemize}

The Galactic environment, although often overlooked, plays a very important role in shaping the planetary system. Our study shows the importance of a homogeneous analysis to reveal the subtle relationships between stellar and planetary properties. We show that both stellar mass and metallicity affect the formation of high and low-mass planets. We provided observational data to investigate the impact of the chemical evolution of the Milky Way on the current planet populations. Future observations and detection of planets in very different environments, such as open and globular clusters, or even dwarf galaxies in the local group \citep{Magrini2023arXiv231208270M}, may expand our knowledge of the processes of formation and evolution of planetary systems, correlating them with stellar and galactic properties. 

\begin{acknowledgements}


We thank the referee for the comments which helped improve the paper. This work has been developed within the framework of the Ariel ``Stellar Characterisation'' working group, in synergy with the ``Planetary Formation'' working group of the ESA Ariel space mission Consortium. L.M. and M.T. thank INAF for the support (Large Grant EPOCH), the Mini-Grants Checs and PILOT (1.05.23.04.02), and the financial support under the National Recovery and Resilience Plan (NRRP), Mission 4, Component 2, Investment 1.1, Call for tender No. 104 published on 2.2.2022 by the Italian Ministry of University and Research (MUR), funded by the European Union –NextGenerationEU– Project ‘Cosmic POT’ Grant Assignment Decree No. 2022X4TM3H by the Italian Ministry of Ministry of University and Research (MUR). The authors acknowledge the support of the ARIEL ASI-INAF agreement n.2021-5-HH.2-2024. 

C.D. acknowledges financial support from the INAF initiative ``IAF Astronomy Fellowships in Italy'', grant name \textit{GExoLife}. 

D.B. acknowledges funding support by the Italian Ministerial Grant PRIN 2022, ``Radiative opacities for astrophysical applications'', no. 2022NEXMP8, CUP C53D23001220006. 

T.L.C. is supported by Funda\c c\~ao para a Ci\^encia e a Tecnologia (FCT) in the form of a work contract (CEECIND/00476/2018).

E.D.M. acknowledges the support from FCT through national funds and from FEDER through COMPETE2020 by the following grants: UIDB/04434/2020, UIDP/04434/2020, and 2022.04416.PTDC. E.D.M. further acknowledges the support from FCT through Stimulus FCT contract 2021.01294.CEECIND. EDM acknowledges the support by the project MICIU/AEI/PID2023-150468NB-I00 and by the Ram\'on y Cajal contract RyC2022-035854-I funded by the Spanish MICIU/AEI/10.13039/501100011033 and by ESF+ .

CPF and HR have received funding from the European Union's Horizon Europe research and innovation programme under grant agreement No. 101079231 (EXOHOST), and from the United Kingdom Research and Innovation (UKRI) Horizon Europe Guarantee Scheme (grant number 10051045). 

KGH is supported by the Polish National Science Center through grant no. 2023/49/B/ST9/01671. Polish participation in SALT is funded by MEiN grant No. 2021/WK/01. 
\\

The team is very grateful to the service astronomers that performed our observations at ESO (with UVES during P105, P106, 109, 110, 111), at the TNG (with HARPS-N during A41, A42) and at the LBT (with PEPSI during 2021). Based on observations collected at the European Southern Observatory under ESO programmes 105.20P2.001, 106.21QS.001, 109.23J9.001, 110.24BU.001 and 111.2542.001; at the LBT observatory under programme 2021\_2022\_25; at the Italian Telescopio Nazionale Galileo (TNG) under programmes AOT41\_TAC25 and AOT42\_TAC20. We acknowledge the use of the archival data from the following ESO programs: 093.C-0417, 094.C-0428, 095.C-0367, 096.C-0417, 097.C-0571, 098.C-0292, 099.C-0374, 0101.C-0407, 099.C-0491, 0100.C-0487, 098.C-0820, 099.C-0303, 0100.C-0474, 0101.C-0497, 097.C-0948, 098.C-0860, 0100.C-0808, 0101.C-0829, 074.C-0364, 60.A-9700(G), 198.C-0169, 096.C-0657, 0102.C- 0618, 0102.C-0319, 191.C-0873. 
\\

The LBT is an international collaboration among institutions in the United States, Italy and Germany. LBT Corporation partners are: Istituto Nazionale di Astrofisica, Italy; The University of Arizona on behalf of the Arizona Board of Regents; LBT Beteiligungsgesellschaft, Germany, representing the Max-Planck Society, The Leibniz Institute for Astrophysics Potsdam, and Heidelberg University; The Ohio State University, representing OSU, University of Notre Dame, University of Minnesota and University of Virginia. The TNG is operated by the Fundación Galileo Galilei (FGG) of the Istituto Nazionale di Astrofisica (INAF) at the Observatorio del Roque de los Muchachos (La Palma, Canary Islands, Spain).\\

This research has made use of the NASA Exoplanet Archive, which is operated by the California Institute of Technology, under contract with the National Aeronautics and Space Administration under the Exoplanet Exploration Program. This work has made use of the VALD database, operated at Uppsala University, the Institute of Astronomy RAS in Moscow, and the University of Vienna. This work presents results from the European Space Agency (ESA) space mission {\em Gaia}. {\em Gaia} data are being processed by the {\em Gaia} Data Processing and Analysis Consortium (DPAC). Funding for the DPAC is provided by national institutions, in particular the institutions participating in the Gaia MultiLateral Agreement (MLA). The {\em Gaia} mission website is https://www.cosmos.esa.int/gaia. The {\em Gaia} archive website is https://archives.esac.esa.int/gaia. This publication makes use of data products from the Two Micron All Sky Survey, which is a joint project of the University of Massachusetts and the Infrared Processing and Analysis Center/California Institute of Technology, funded by the National Aeronautics and Space Administration and the National Science Foundation.
\\

We used the python packages: astropy \citep[\url{http://www.astropy.org},][]{Astropy2022}, numpy \citep[\url{https://numpy.org},][]{harris2020}, PyAstronomy \citep[\url{https://pyastronomy.readthedocs.io/en/latest/index.html},][]{pyastronomy2019}, scipy \citep[\url{https://www.scipy.org},][]{2020SciPy-NMeth}, pandas \citep[\url{https://pandas.pydata.org},][]{mckinney-proc-scipy-2010}, and matplotlib \citep[\url{https://matplotlib.org},][]{Hunter2007}. \\

\end{acknowledgements}

\bibliography{bibliography} 

\appendix

\section{Stellar parameters}\label{sample_params}

In the online Table~\ref{sample_params}, we provide the stellar sample presented by their coordinates (right ascension and declination), name, the effective temperature (\teff), surface gravity (\logg), metallicity (\feh), microturbulent velocity (\vt), macroturbulent velocity (\vmac), projected rotational velocity (\vsini), stellar mass (\Mstar) with their uncertainties. We provide a flag for the microturbulent velocity: 0 indicates that \vt~ is derived from spectral analysis, and 1 that it is assumed from the relation between \vt~ and \teff, \logg, and \feh. The last column, Spectra, shows where the spectra were taken.

\begin{figure*}
  \centering
  \includegraphics[width=0.8\linewidth]{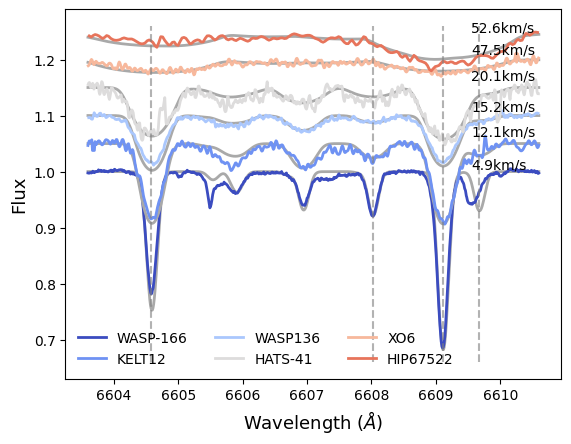}  
  \caption{Six different spectra of stars analysed with the spectra synthesis technique for a segment of the wavelength region. The \vsini~ is indicated in the plot. The dashed horizontal lines represent the iron lines in this region.}
  \label{spectra}
\end{figure*}

\begin{figure*}
  \centering
  \includegraphics[width=0.33\linewidth]{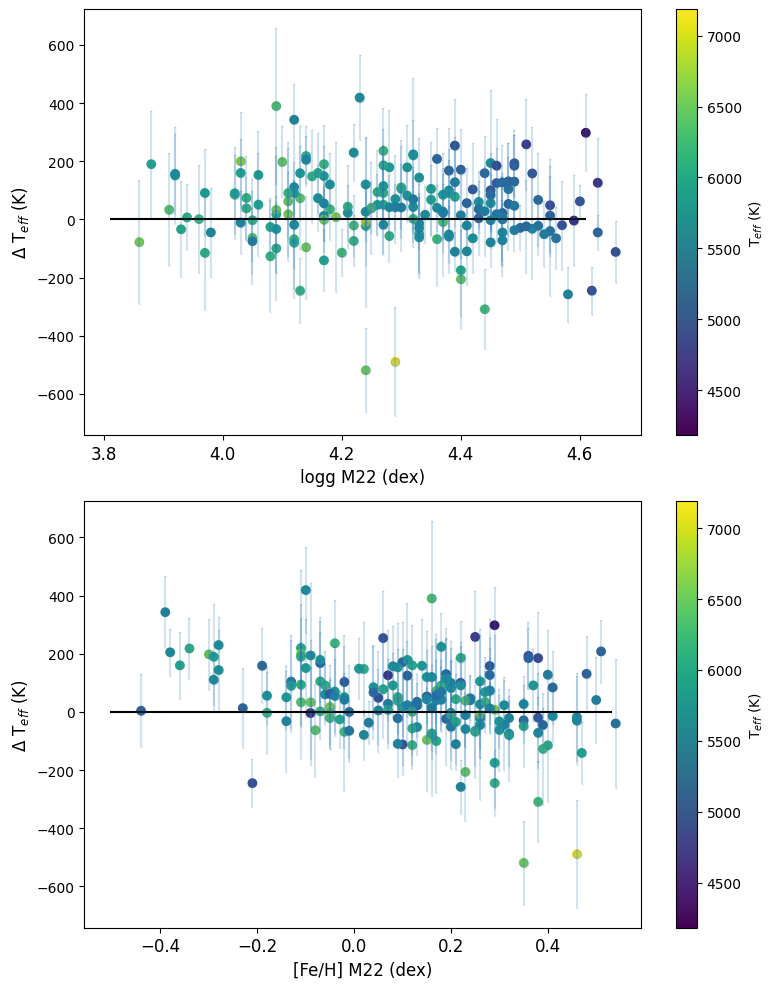}   
  \includegraphics[width=0.33\linewidth]{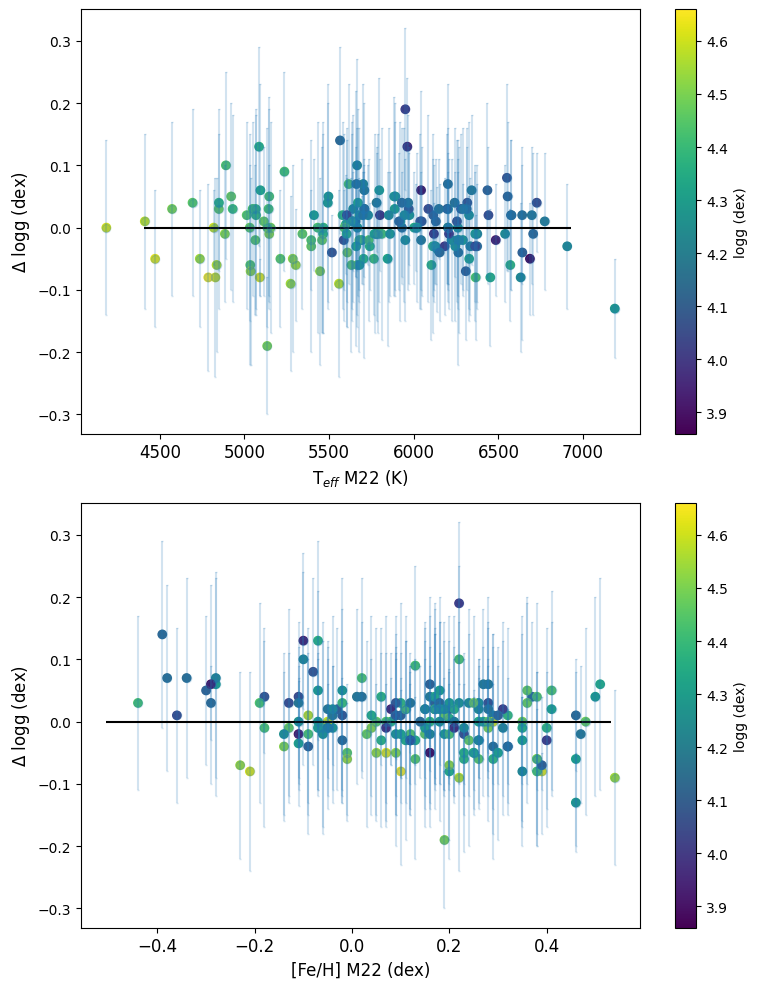}   
  \includegraphics[width=0.33\linewidth]{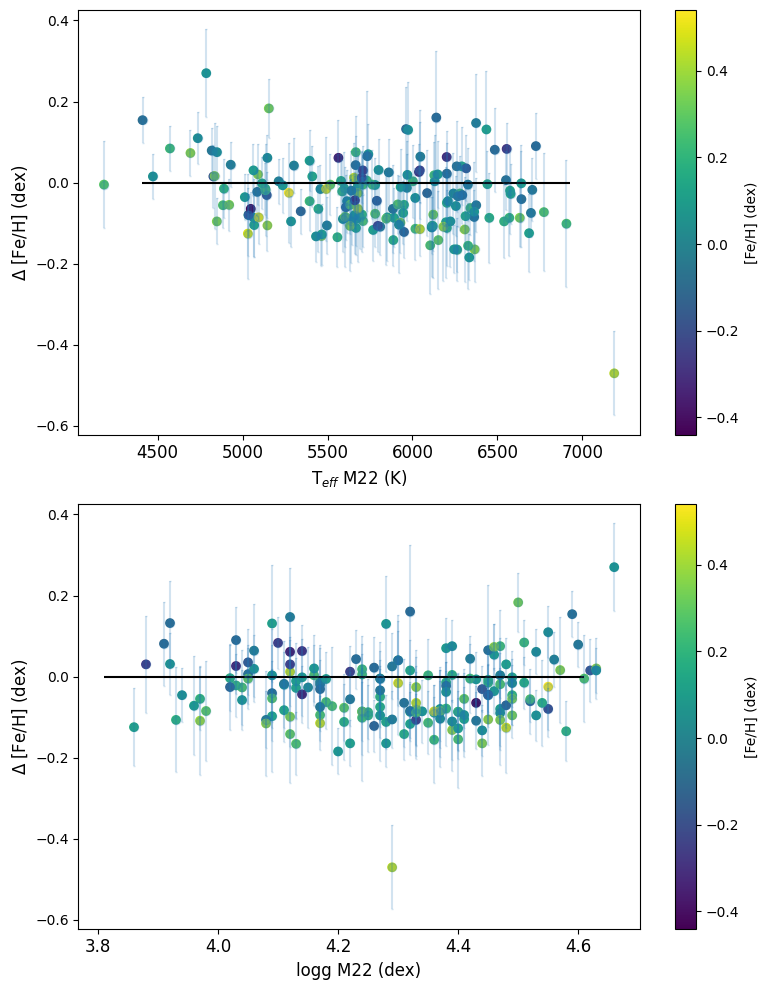}  
  \caption{Results of the spectral synthesis analysis in comparison with those of M22. The panels show the differences (this work–M22) in \teff\, (left panel), \logg\, (central panel) and \feh\, (right panel), respectively, as a function of the parameters derived in M22.}
  \label{param_dependence}
\end{figure*}

\begin{figure}
  \centering
  \includegraphics[width=0.95\linewidth]{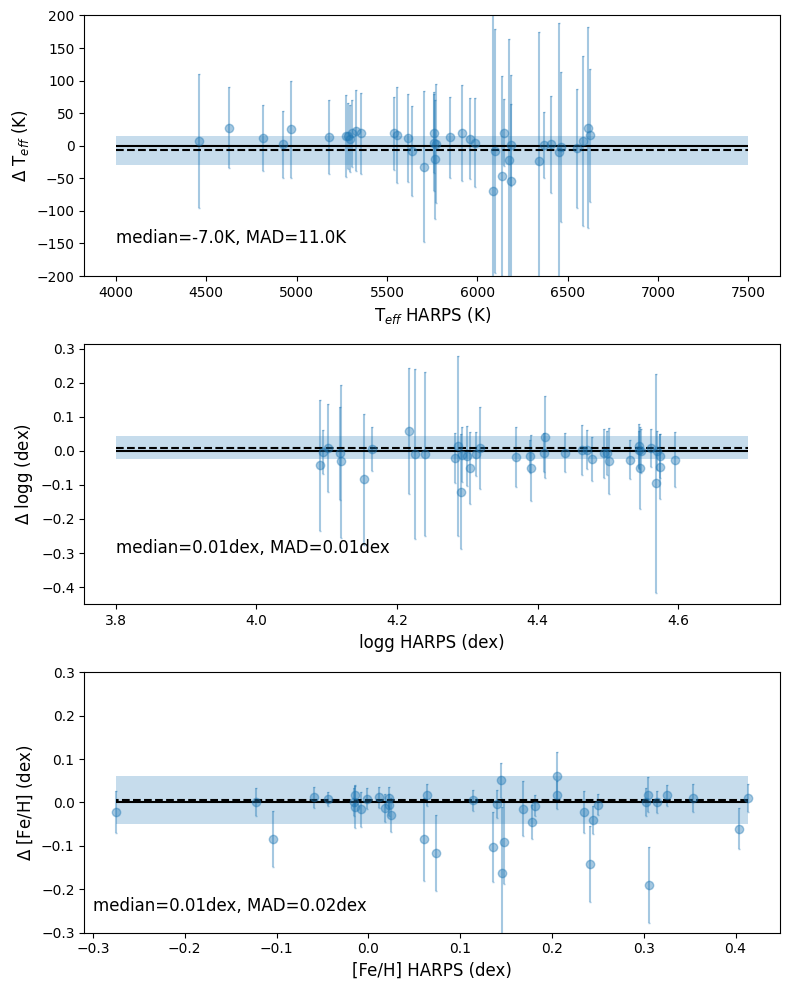}   
  \caption{Comparison of stellar parameters between a sample of the HARPS spectra with spectra convolved to the resolution of 48\,000 from a purely spectroscopic analysis (Run 0). The panels show the differences (HARPS--Convolved) in \teff~ (upper panel), \logg$_{\rm trig}$ (central panel) and \feh~ (lower panel) as a function of the respectively parameters. The shaded areas correspond to the 1\,$\sigma$ of the differences from the median. The dashed lines show the median differences for each parameter while the solid lines are zero.}
  \label{comparison_resolution}
\end{figure}

\section{Projected rotational velocities for the Ariel planet hosts}\label{vsini_ariel}

In the online Table~\ref{vsini_ariel}, we provide the projected rotational velocity (\vsini) and their uncertainties for the stars in this work and M22. 

\section{Orbital properties}\label{orbital}

In the online Table~\ref{orbital}, we provide U, V, W velocities, galactocentric distance, and orbital eccentricity. 

\section{Planetary properties}\label{app:planet_params}

\begin{figure}
\centering
  \includegraphics[width=1\linewidth]{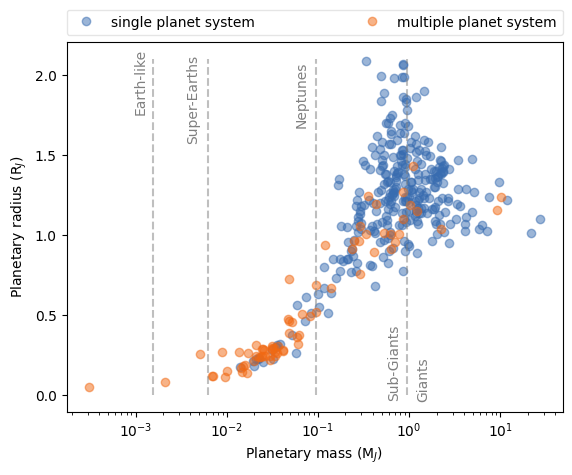}  
  \caption{Masses and radii of planets orbiting the 353 stars in the Ariel MCS taken from \cite{Edwards2019, Edwards2022} which are based on the NASA exoplanet catalogue. The different mass categories for the planets are indicated with the gray dashed lines and are taken from \cite{Emsenhuber2021}.}
\label{planet_param_fig}
\end{figure}

\begin{figure}
  \centering
  \includegraphics[width=0.95\linewidth]{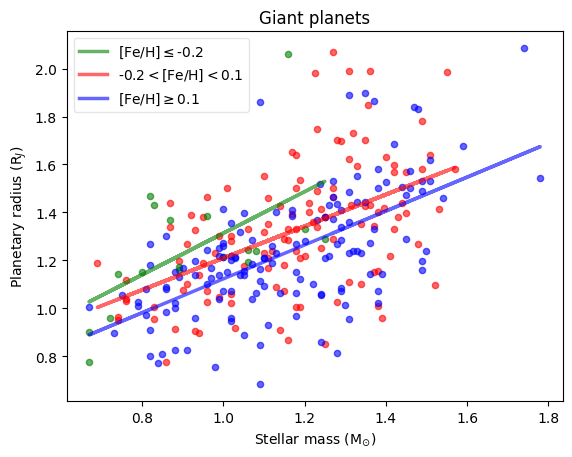}
  \caption{Planet radius of giant planets in the sample as a function of stellar mass derived in this work using the metallicity bins of M22. The super-solar metallicities (\feh\,$\geq$\,0.1\,dex) are shown in blue, the solar metallicities (--0.2\,$<$\,\feh\,$<$\,0.1\,dex) in red, and the sub-solar metallicities (\feh\,$\leq$--0.2\,dex) in green. The linear fits to the three datasets are colour-coded in the same way.}
  \label{fig:stellarmass_planetradius_giants}
\end{figure}

\begin{figure}
  \centering
  \includegraphics[width=0.95\linewidth]{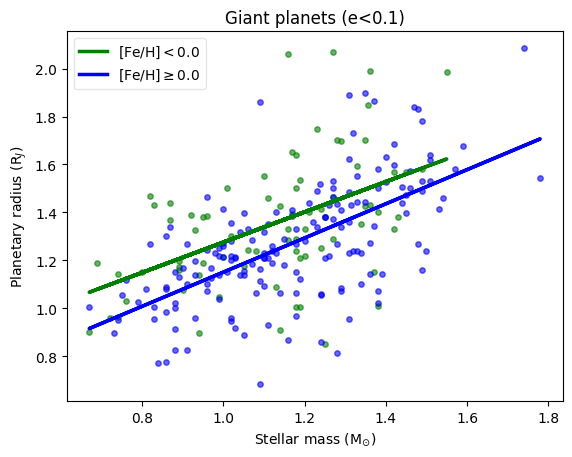}
  \caption{Planetary radius as a function of homogeneous stellar mass for the single-planet systems hosting a giant same as in Fig.~\ref{fig:stellarmass_planetradius_giants_2bins} but only for systems with eccentricity $e < 0.1$. The planets in metal-rich hosts (\feh$\geq$ 0.0 dex) are shown in blue and the metal poor (\feh$<$ 0.0 dex) in green. The linear fits to the two datasets are colour-coded in the same way.}
  \label{fig:R_M_feH_elt01}
\end{figure}

\begin{figure}
  \centering
  \includegraphics[width=0.95\linewidth]{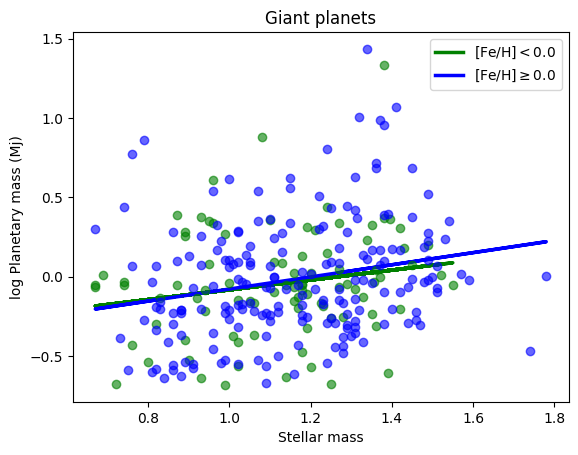}
  \caption{Planetary mass as a function of homogeneous stellar mass for the single-planet systems hosting a giant. The planets in metal-rich hosts (\feh$\geq$ 0.0 dex) are shown in blue and the metal poor (\feh$<$ 0.0 dex) in green. The linear fits to the two datasets are colour-coded in the same way.}
  \label{fig:stellarmass_planetmass_giants}
\end{figure}

\end{document}